\documentclass[10pt]{article}
\pdfoutput=1
\usepackage{graphicx}
\usepackage{epsfig}
\usepackage{bbm}
\usepackage{cite}

\newcommand{\be}{\begin{eqnarray}}
\newcommand{\ee}{\end{eqnarray}}

\def\nn{\nonumber}

\def\gs{\mathrel{
   \rlap{\raise 0.511ex \hbox{$>$}}{\lower 0.511ex \hbox{$\sim$}}}}
\def\ls{\mathrel{
   \rlap{\raise 0.511ex \hbox{$<$}}{\lower 0.511ex \hbox{$\sim$}}}}
\newcommand{\bea}{\begin{equation} \begin{array}{c}}
\newcommand{\bead}{\begin{equation} \begin{array}{cccc}}
\newcommand{\eea}{ \end{array} \end{equation}}

\newcommand{\obb}{0\mbox{$\nu\beta\beta$}}
\newcommand{\onbb}{neutrinoless double beta decay}

\begin{document}

\title{\bf Lepton Number and Lepton Flavor Violation Through Color Octet States
}
\author{
Sandhya Choubey,$^{a,}$\thanks{email: \tt sandhya@hri.res.in}~~
Michael Duerr,$^{b,}$\thanks{email: \tt michael.duerr@mpi-hd.mpg.de}~~
Manimala Mitra,$^{c,}$\thanks{email: \tt  manimala.mitra@lngs.infn.it}~~ 
Werner Rodejohann$^{b,}$\thanks{email: \tt werner.rodejohann@mpi-hd.mpg.de}
\\\\
\\
{\normalsize \it$^{a}$Harish-Chandra Research Institute, Chhatnag Road, Jhunsi, 211019 Allahabad, India}\\ \\ 
{\normalsize \it$^{b}$Max-Planck-Institut f\"ur Kernphysik, Postfach
103980, 69029 Heidelberg, Germany }\\ \\
{\normalsize \it$^{c}$I.N.F.N., Laboratori Nazionali del Gran Sasso, 
67010 Assergi (AQ), L'Aquila, Italy}
}
\date{}

\maketitle
\begin{abstract}\noindent
We discuss neutrinoless double beta decay and lepton flavor violating
decays such as $\mu \to e\gamma$ in the colored seesaw scenario. In
this mechanism, neutrino masses are generated at  one-loop via the exchange 
of TeV-scale fermionic and scalar color octets. The same
particles mediate lepton number and flavor violating processes. 
We show that within this framework a dominant color octet contribution to 
neutrinoless double beta decay is possible without being in conflict 
with constraints from lepton flavor violating processes. We
furthermore compare the ``direct'' color octet contribution to neutrinoless
 double beta
decay with the ``indirect'' contribution, namely the usual standard
light Majorana neutrino exchange. For degenerate color octet fermionic
states both contributions are proportional to the usual effective mass, while for
non-degenerate octet fermions this feature is not present.  Depending
on the model parameters, either of the contributions can be dominant.

\end{abstract}

\newpage 

\section{Introduction}
The question regarding the origin of neutrino mass remains one of
the most pressing problems in particle physics \cite{revs}. 
In particular, knowledge of whether neutrinos are Dirac or Majorana particles 
is of paramount importance. If neutrinos were to be Majorana particles, 
lepton number would be violated. Since lepton number is an accidental symmetry of the standard model of particle 
physics, one would like 
to know the
scale of lepton number violation (LNV), if any,
leading to the generation of neutrino masses. In addition, the peculiar pattern of neutrino masses and mixing 
demands an explanation which requires the 
extension of the standard model, and one would like to probe the 
existence of  new particles usually considered within such extensions.  
While neutrino oscillations and beta decay experiments continue to 
improve our knowledge of the low energy neutrino mass matrix, 
we have to look at data from a variety of complementary avenues 
such as neutrinoless double beta decay (\obb) experiments  \cite{epj, cuoricino+nemo, Gerda, Cuore}, 
lepton flavor violation searches \cite{MEG, belle}, as well as
collider experiments to augment our understanding of the mechanism responsible for neutrino mass generation. 
On the experimental side, there are realistic prospects for order-of-magnitude 
improvements in the search for the $0\nu\beta\beta$ 
half-life \cite{future}. Moreover, there has been an order-of-magnitude
improvement on the bound for $\mu \to e \gamma$ recently \cite{MEG}.

While one usually assumes that the exchange of light Majorana neutrinos is
the leading contribution to neutrinoless double beta 
decay \cite{0nu2beta-old}, it is well known that a
plethora of alternative intermediate scale 
theories exists (where the intermediate scale is TeV, few hundred GeV and even few tens of GeV), 
which not only predict neutrinoless double 
beta decay, but can  also 
saturate the current bounds on the
process, or can lead to sizable rates in current and future experiments
\cite{feinberg, ms0nu2beta, mwex, pion-ex, hirsch,  allanach, vogel, choi,  tello, Ibarra, blennow, msv, santamaria}. 
Some  of the interesting alternative approaches are based 
on seesaw models
\cite{Ibarra, blennow, msv}, 
left-right symmetric theories \cite{tello},  $R$-parity violating supersymmetry \cite{ms0nu2beta, mwex,
pion-ex,hirsch, allanach} and so on. See the
recent reviews \cite{Werner-rev} and \cite{ital-rev} for summaries of
the particle physics aspects and the experimental situation of \obb,
respectively. 
The fact that heavy TeV-scale particles can lead to the same contribution to
\obb~as sub-eV scale neutrinos is easily understood from looking at the
particle physics amplitude in the standard interpretation of light
neutrino exchange: 
\be
{\cal A}_l \simeq G_F^2 \, \left| \sum_i \frac{U_{ei}^2 \, m_i}{\langle p^2
\rangle - m_i^2} \right|
\simeq \left( 2.7 \, {\rm TeV}\right)^{-5} \, . 
\ee
Here we have inserted the current limit of about 0.5 eV on the
effective neutrino mass $M_{ee} = |\sum U_{ei}^2 \, m_i|$, with $U$ being the lepton
mixing matrix, $m_i$ the neutrino masses,  and $ \langle p^2
\rangle \simeq 0.01$ GeV$^{2} \gg m_i^2$ for the relevant momentum scale of the
process in the neutrino propagator. Thus, if heavy particles with mass
much larger than $ \langle p^2\rangle$ are exchanged in the process,
for instance two scalars or vectors with mass $M_1$ and one fermion
with mass $M_2$,
they will contribute with $\sim 1/(M_1^4 \, M_2)$ to the
amplitude, and can saturate the current limit if their masses are in
the TeV regime. 

In this paper we work with 
a particular variant of the seesaw mechanism \cite{seesaw} 
which has TeV-scale scalar and  
fermionic color octets. The fermionic octets have a Majorana mass 
term leading to lepton number violation in the theory. Consequently, one expects that they will directly generate \onbb. 
Neutrino 
masses are forbidden at tree level, however they are generated at one-loop level, via 
the exchange of  color octet fermions and scalars. 
This model has been dubbed the "colored seesaw" model 
\cite{FileviezPerez:2009ud}. Since the new particles introduced in this 
model are color octets, sizable cross-sections 
at the LHC can be expected, connected with the
spectacular feature of lepton number violation
\cite{FileviezPerez:2010ch}. In addition, the lepton flavor violation
associated with the gauge invariant Yukawa couplings of the octets 
\cite{Liao:2009fm}
implies that not only the neutrino sector is non-trivial in what
regards flavor, but also the charged lepton sector is. Hence, decays like
$\mu \to e \gamma$ provide additional tests of the scenario \cite{Liao:2009fm}, and the 
branching ratios depend on the same set of parameters as the
neutrino mass matrix. 
Here, we discuss neutrinoless double beta decay and lepton flavor violation mediated 
by these color octet states, and show that they, being TeV-scale particles, can give a large and in fact saturating 
  contribution in both processes.
Note that 
the current collider limits on their masses are 
 1.92 TeV for the octet scalar \cite{LHC}, and about 1 TeV for the octet
fermions \cite{LHC1}.

The reason why these contributions have not been considered in the
literature so far is that usually one assumes minimal flavor
violation (MFV) 
to avoid large flavor changing neutral current (FCNC) effects 
\cite{Manohar:2006ga}. Then, the coupling of the charged member in the weak doublet scalar
octet to a quark $q$ is proportional to $m_{q}/v$, with $v$ being the  vacuum 
expectation value (VEV) of the standard
model Higgs. Therefore, the amplitude of \obb~receives a suppression
factor of $m_{u,d}^2/v^2$ and is completely negligible
\cite{FileviezPerez:2009ud}. We depart from the  assumption of  
 minimal flavor violation in this work, and
hence  a sizable rate of
\obb~can come from the color octet states. 
In this framework, we study the
interplay of neutrino mass and mixing, lepton flavor violation and
neutrinoless double beta decay, 
and encounter
several interesting and general features, which are characteristic for
the correlations between the various phenomenological sectors. 

One particularly interesting feature is the interplay of direct and
indirect contributions of the color octets to \obb. With ``direct 
contribution'' we mean the new contribution with exchange of heavy
octets, while with ``indirect contribution'' we mean the usually
considered light Majorana neutrino exchange
mechanism, where here, however, the neutrino masses are generated by the octets via
loops. While in general these two contributions depend differently on
the parameters, 
we identify situations in which they are
both proportional to the effective mass, namely when the fermionic
octets are degenerate in mass. 
In other cases, they depend differently
on the flavor parameters. In all cases, the neutrino exchange mechanism can be
dominating or sub-leading, depending on the masses of the octets and
the quartic coupling governing the interaction between the color octet scalar and the standard model Higgs boson.

The outline of the paper is the following: we start with the basics of the 
colored seesaw scenario in Section \ref{cols}. Following this, in 
Section \ref{0nu2beta-lfv}, we discuss the contribution of color octet
states in neutrinoless double beta decay and lepton flavor 
violating processes in general, and consider in some detail the most simple case of two
degenerate color octet fermions in Section \ref{twodeg}. The general
case of three non-degenerate color octet fermions is the subject of
Section \ref{threegen}, before we conclude and summarize in Section \ref{sum}.

\begin{figure}
 \centering
\includegraphics[scale=1]{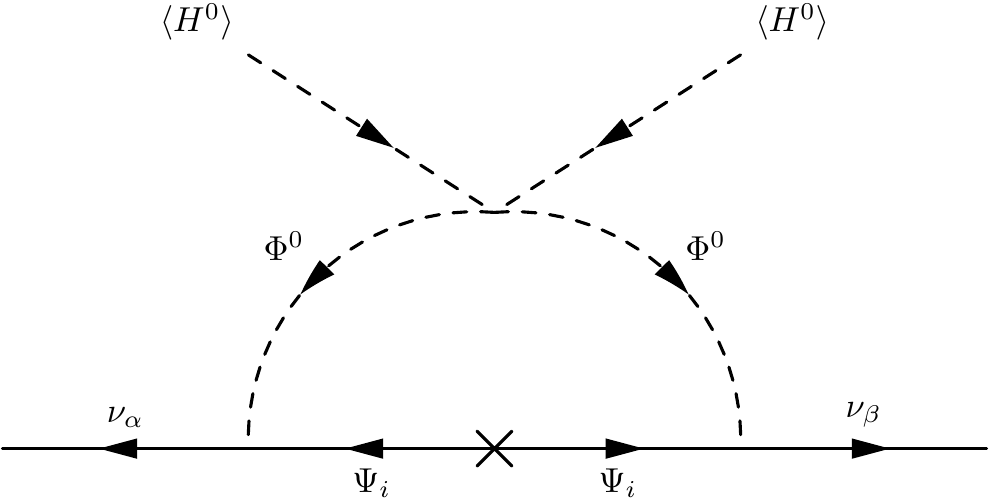}
\caption{Neutrino mass generated at the one-loop level in the colored seesaw scenario.}\label{neumass}
\end{figure}

\section{ Colored Seesaw Scenario \label{cols}}

In the colored seesaw mechanism \cite{FileviezPerez:2009ud,FileviezPerez:2010ch}, the particle content of the standard model is extended 
with a color octet scalar $\Phi$ and color octet fermions $\Psi_i$, which transform as
\be
\Phi \sim (8,2,+1)~;~~ \Psi_i \sim (8,1,0) \, .
\ee
While $\Phi$ is 
charged under $SU(2)_L$ and $U(1)_Y$, the fields $\Psi_i$ are singlets
under these gauge groups. Note that while the model just contains one additional scalar, 
we need at least two additional fermions in order to generate at least two massive light neutrinos which are required to explain current neutrino data. 
We will consider the cases $i = 1,2$ and $i=1,2,3$ in this paper. 
The relevant Lagrangian corresponding to the new sector is  
\be \label{eq:lagrangian-nus}
-{ \it \mathcal{L}}_{\nu}=Y_{\nu}^{\alpha i} \bar{L}_\alpha \,
i\sigma_2 \, \mathrm{Tr}(\Phi^\ast \Psi_i)+ \frac{1}{2}M_{\Psi_i} 
\mathrm{Tr}( \bar{\Psi}_i^c \Psi_i)+\lambda_{\Phi H} 
\mathrm{Tr}(\Phi^{\dagger} H)^2+ \rm{h.c.},
\ee
where $H$ is the standard model doublet Higgs and 
$L_\alpha$ is the lepton doublet of flavor $\alpha = e, \mu,
\tau$, i.e., greek indices correspond to flavor and roman indices to
physical mass states.  We have considered $\lambda_{\Phi H}$ and $M_{\Psi_i}$ 
as real.
Since the scalar $\Phi$ is a color octet and $SU(3)_c$ symmetry must remain 
unbroken, it must have a zero vacuum expectation value. Hence, it is
evident  from the above that at tree level one cannot generate the
neutrino mass operator $\mathcal{O} (\frac{LH LH}{\Lambda})$ \cite{dim5}. 
However,
neutrino masses will be generated at one-loop level through the
mediation of colored octet scalars and fermions, as shown in
\cite{FileviezPerez:2009ud,FileviezPerez:2010ch} and illustrated in Fig.~\ref{neumass}. 
The one-loop neutrino mass matrix is 
\be \label{eq:numass}
{M}^{\alpha \beta}_\nu= \sum_i v^2 \frac{\lambda_{\Phi H}} {16 \pi^2}
\, Y^{ \alpha i}_{\nu} Y^{\beta i}_{\nu} \, \mathcal{I}(M_{\Phi},
M_{\Psi_{i}}) \, ,
\label{mnu}
\ee
where the loop function $\mathcal{I}$ is given by
\be
\mathcal{I}_i \equiv 
\mathcal{I}(M_{\Phi}, M_{\Psi_i})=M_{\Psi_i}
\frac{M^2_{\Phi}-M^2_{\Psi_i}+ M^2_{\Psi_i} \ln(\frac
{M^2_{\Psi_i}}{M^2_{\Phi}})}{ (M^2_{\Phi}-M^2_{\Psi_i})^2}  .
\ee
Here, $M_\Phi$ is the mass of the scalar octet, and
$v=174\,\mathrm{GeV}$ is the doublet Higgs VEV. 
The neutrino mass matrix in Eq.~(\ref{mnu}) can be diagonalized by the
unitary  $3 \times 3$ Pontecorvo--Maki--Nakagawa--Sakata (PMNS)  mixing matrix $U$, where 
according to our convention $ U^{\dagger} M_\nu U^\ast=
M^d_\nu = {\rm diag}(m_1, m_2, m_3)$, with $m_i$ being the light neutrino
masses. Using this and Eq.~(\ref{mnu}), one can express the Yukawa
coupling matrix $Y_{\nu}$ in terms of light neutrino masses, mixings, color octet
masses and a Casas--Ibarra \cite{casas-Ibarra} matrix $\mathcal{R}$ as: 
\be
Y_{\nu}=\sqrt{\frac{16 \pi^2}{\lambda_{\Phi H} }}\frac{1}{v} \, U \,
\sqrt{M^d_\nu} \, 
\mathcal{R} \,  \sqrt{(\mathcal{I}^d)^{-1}} \, ,
\ee
where $\mathcal{I}^d=\rm{diag}
(\mathcal{I}_1,\mathcal{I}_2,\mathcal{I}_3 )$. In general, we can
express the complex orthogonal matrix $\mathcal{R}$ as 
\be
\mathcal{R}=\pmatrix{ \hat{c}_2 \hat{c}_3 & \hat{c}_2 \hat{s}_3 & \hat{s}_2 \cr 
-\hat{c}_1 \hat{s}_3-\hat{s}_1 \hat{s}_2 \hat{c}_3 & \hat{c}_1\hat{c}_3-\hat{s}_1\hat{s}_2\hat{s}_3 & \hat{s}_1\hat{c}_2 \cr
\hat{s}_1\hat{s}_3-\hat{c}_1\hat{s}_2\hat{c}_3 &
-\hat{s}_1\hat{c}_3-\hat{c}_1\hat{s}_2\hat{s}_3 & \hat{c}_1\hat{c}_2} ,
\label{rfor3}
\ee
with $\hat{s}_i = \sin \hat{\theta}_i$, $\hat{c}_i = \cos \hat{\theta}_i$ 
($i=1,2,3$), where $\hat{\theta}_1$, $\hat{\theta}_2$ and
$\hat{\theta}_3$ are arbitrary complex angles. 
Since the color octet scalar field $\Phi$ is 
charged under $SU(3)_c$ as well as $SU(2)_L$ and $U(1)_Y$, it can interact with 
the quark fields of the standard model. Note that the scalar field $\Phi$ can be decomposed as
\be
\Phi= \lambda^A \pmatrix{ \Phi^+ \cr \Phi^0}_A~;~~ \Phi^0=\Phi^0_r + i \Phi^0_i \, ,
\ee
where the $\lambda^A$, $A = 1, \ldots, 8$, are the Gell-Mann matrices. 
The Lagrangian corresponding to the interaction between 
the quarks and the scalar $\Phi$   is 
\be \label{eq:lagrangian-quarks}
{\it \mathcal{L}}_Q= \bar{d}_R \kappa_D \Phi^{\dagger} Q_L+\bar{u}_R \kappa_U Q_L \Phi+\rm{h.c.} ,
\ee
where $\kappa_{U,D}$ are the corresponding Yukawa couplings.  Going
from the flavor basis to the  physical basis of 
the quark fields ($U_{L,R}$ and $D_{L,R}$ are the bases rotation matrices for the up and down quarks, respectively), the 
Lagrangian is 
\be
\mathcal L_Y=
&\bar d\left[P_L\left(D_R^\dagger \kappa_DU_L\right)-P_R\left(D_L^\dagger \kappa_U^\dagger U_R\right)\right]\Phi^- u+\bar u\left[P_R\left(U_L^\dagger \kappa_D^\dagger D_R\right)-P_L\left(U_R^\dagger \kappa_U D_L\right)\right]\Phi^+d \nn
\\ 
&+\frac{\Phi_r^0}{\sqrt2}\bar d\left[P_L\left(D_R^\dagger \kappa_D D_L\right)+P_R\left(D_L^\dagger \kappa_D^\dagger D_R\right)\right]d+\frac{\Phi_r^0}{\sqrt2}\bar u\left[P_L\left(U_R^\dagger \kappa_U U_L\right)+P_R\left(U_L^\dagger \kappa_U^\dagger U_R\right)\right]u
 \\
&-i\frac{\Phi_i^0}{\sqrt2}\bar d\left[P_L\left(D_R^\dagger
\kappa_DD_L\right) - P_R\left(D_L^\dagger \kappa_D^\dagger
D_R\right)\right]d+i\frac{\Phi_i^0}{\sqrt2}\bar
u\left[P_L\left(U_R^\dagger \kappa_U U_L\right)-P_R\left(U_L^\dagger
\kappa_U^\dagger U_R\right)\right]u \, .\nn
\ee
We choose the basis where the up-quark mass matrix is diagonal, i.e.,
the up-quark mixing matrices are $U_{L,R}=\mathbbm{1}$. Using this, the  above equation
simplifies to 
\be
\mathcal L_Y=
&\bar d\left[P_L\left(D_R^\dagger \kappa_D\right)-P_R\left(D_L^\dagger \kappa_U^\dagger \right)\right]\Phi^-u+\bar u\left[P_R\left(\kappa_D^\dagger D_R\right)-P_L\left( \kappa_U D_L\right)\right]\Phi^+d \nn
\\
&+\frac{\Phi_r^0}{\sqrt2}\bar d\left[P_L\left(D_R^\dagger \kappa_D D_L\right)+P_R\left(D_L^\dagger \kappa_D^\dagger D_R\right)\right]d+\frac{\Phi_r^0}{\sqrt2}\bar u\left[P_L\left( \kappa_U \right)+P_R\left( \kappa_U^\dagger \right)\right]u
\\ 
&-i\frac{\Phi_i^0}{\sqrt2}\bar d\left[P_L\left(D_R^\dagger
\kappa_DD_L\right) - P_R\left(D_L^\dagger \kappa_D^\dagger
D_R\right)\right]d+i\frac{\Phi_i^0}{\sqrt2}\bar u\left[P_L\left(
\kappa_U \right)-P_R\left( \kappa_U^\dagger \right)\right]u \nn \, .
\ee
After describing the basic Lagrangian in this section, we now discuss
the different processes where these color octet states can
participate. In particular, we concentrate on the novel color octet contribution
to neutrinoless double beta decay and on lepton flavor violating processes like $\mu \to
e\gamma$. We will show in particular that the color octet framework
for neutrino mass generation can easily saturate both the $\mu \to e
\gamma$ limit as well as the limit on \onbb.

\section{Neutrinoless Double Beta Decay and Flavor Violation \label{0nu2beta-lfv}}

\begin{figure}
\centering
\begin{minipage}{0.4\linewidth}
\centering
\includegraphics[scale=1]{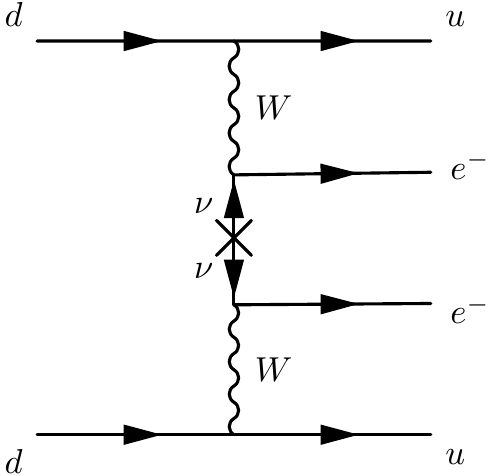}
\caption{Neutrinoless double beta decay mediated by light Majorana neutrinos.}\label{0nu2beta-nus}
\end{minipage}
\hspace{.1\linewidth}
\begin{minipage}{0.4\linewidth}
\centering
\includegraphics[scale=1]{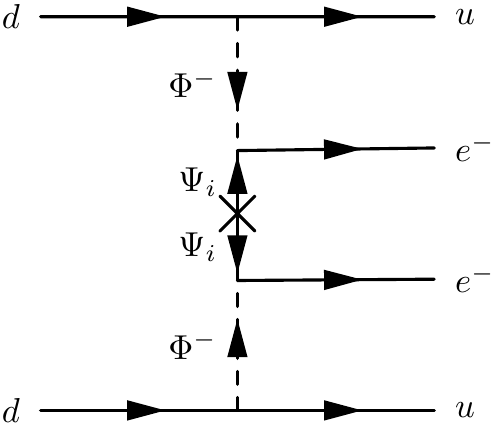}
\caption{Neutrinoless double beta decay mediated by color octet scalars and fermions.}\label{0nu2beta-color}
\end{minipage}
\end{figure}

The standard light neutrino contribution to \obb~is given in 
Fig.~\ref{0nu2beta-nus}. It is proportional to the effective mass
$M_{ee} = \sum U_{ei}^2 m_i$. We note in this paper that the scalar and fermionic states
will also contribute in this lepton number violating process, as shown
in Fig.~\ref{0nu2beta-color}. The conceptually interesting part is
that there is a direct and an indirect contribution from the
octets. The direct one is the new diagram from
Fig.~\ref{0nu2beta-color}, while the indirect one is the neutrino
exchange mechanism in Fig.~\ref{0nu2beta-nus}. It is called indirect
because the light neutrino masses are generated by the octets in the loop
diagram given in Fig.~\ref{neumass}. The relevant vertices  for 
the direct contribution of the octets to $0\nu \beta\beta$ are 
\be
\lambda^A_{\alpha \beta} \bar{u}^{\alpha} \left[P_R\left(\kappa_D^\dagger D_R\right)_{11}-P_L\left( \kappa_U D_L\right)_{11}\right] 
\Phi^+_Ad^{\beta} \mbox{ and }  Y^{ei}_{\nu}\bar{e}_L (\Phi^+)^* \Psi_i \, , 
\ee 
so that the relevant effective operator for $0\nu \beta\beta$ is $\langle  uuee |{\mathcal L}^{{{\Delta L}_e}=2}_{\rm eff}(x)| dd \rangle$, 
where 
${\mathcal L}^{{{\Delta L}_e}=2}_{\rm eff}(x)$ is 
\be
(Y_{\nu}^{ei})^2 \frac{1}{M^4_{\Phi} M_{\Psi}}\lambda_{\alpha \beta} \lambda_{\gamma \delta} \left[ \bar{u}^{\alpha}\left (P_R
(\kappa_D^\dagger D_R)_{11}-P_L( \kappa_U D_L)_{11}\right )d^{\beta} \right]
\left[ \bar{u}^{\gamma} \left( P_R (\kappa_D^\dagger
D_R)_{11}-P_L(\kappa_U D_L)_{11}\right) d^{\delta} \right] (\bar{e}_L e_L^c)
\, .
\ee
For simplicity, and for illustration, we consider the case where $\kappa_U \ll \kappa_D$ and concentrate on the right-chiral part
only. We denote the contribution coming from the quark 
states as $\tilde{y}^2_{11}=y^2_{11}\beta $, where $y_{11} =
(\kappa_D^\dagger D_R)_{11}$,  and the factor $\beta$ can come from the hadronization procedure. 
The inverse half-life of $0\nu\beta\beta$ is given by the well-known expression  
\be
\frac{1}{T_{1/2}}=G_{0 \nu} |\mathcal{M}_{\nu} \eta_{\nu}+
\mathcal{M}_{\Phi \Psi} \eta_{\Phi \Psi}|^2 \, ,
\mbox{ where } 
\eta_{\nu} = \frac{\sum_i U^2_{ei} m_i}{m_e} \,\mbox{ and } 
\eta_{\Phi \Psi} = \frac{m_p}{G^2_F}\frac{
\tilde{y}^2_{11}}{M^4_{\Phi}}\sum_i\frac{{(Y_{\nu}^{ei})^2}}{
M_{\Psi_i}} \, .  
\label{tp}
\ee
Here $\mathcal{M}_{\nu}$ and $\mathcal{M}_{\Phi \Psi}$ 
are the nuclear matrix elements corresponding to the light neutrino exchange (indirect contribution)
and to the exchange of color octets (direct contribution). We will focus here
for definiteness on the neutrinoless double beta decay of $^{76}$Ge,
for which a half-life limit of $T_{1/2} \geq 1.9 \times 10^{25} \,
\rm{yr}$ exists \cite{epj}. The phase space factor is $G_{0 \nu}=7.93
\times 10^{-15} \, \rm{yr}^{-1}$. Alternatively, one can give the above expression as: 
\be
\frac{1}{T_{1/2}} = \mathcal{K}_{0 \nu} \left|
\frac{{M}_{ee}}{P^2}+\frac{ \tilde{y}^2_{11} }{M^4_{\Phi}
G^2_F}\sum_i\frac{{(Y_{\nu}^{ei})^2}}{ M_{\Psi_i}} \right |^2 \, .
\label{alt}
\ee
In the above, $\mathcal{K}_{0 \nu}=G_{0 \nu}(\mathcal{M}_{\Phi \Psi} 
m_p)^2$,  $P^2 = m_e m_p \frac{ \mathcal{M}_{\Phi
\Psi}}{\mathcal{M}_{\nu}}$ and $G_F = 1.166\times 10^{-5} \, \mathrm{GeV}^{-2}$
is the Fermi constant. The parameter $P^2$ describes the difference
between the long- and short-range contributions. 

It is also instructive
to compare the amplitudes for neutrinoless double beta decay on the {\it particle
physics level}: 
\be \label{amp}
\mathcal{A}_l \simeq G^2_F \frac{{M}_{ee}}{\langle p^2 \rangle} \nn ~,~~
\mathcal{A} \simeq \frac{ y^2_{11}}{M^4_{\Phi} }\sum_i\frac{{(Y_{\nu}^{ei})^2}}{ M_{\Psi_i}} \, .
\ee
Here $\mathcal{A}$ is the amplitude for the direct contribution (color octet exchange)
and $\mathcal{A}_l$ the indirect contribution due to the light Majorana
neutrino exchange.  We deal with a nuclear physics 
problem here, so that a typical size of
the momentum is $\langle p^2 \rangle \approx
(100\,\mathrm{MeV})^2$. 
Comparing different amplitudes on the 
particle physics level is very often an excellent approximation to
compare their relative contribution to \obb. With the help of the complex
orthogonal matrix $\mathcal{R}$ one can re-write the color octet
amplitude in terms of the PMNS matrix: 
\be
\mathcal{A} \simeq \frac{16 \pi^2}{\lambda_{\Phi H} v^2} \frac{y^2_{11}}{M^4_{\Phi}} \left( \sum_{i} m_i U^2_{ei} \sum_j 
\frac{{\mathcal{R}^2_{ij}}}{M_{\Psi_j} \mathcal{I}^d_j}+ 2 
\sum_{j<i} \sqrt{m_i m_j} \, U_{ei} U_{ej} \sum_k \frac{\mathcal{R}_{ik} \mathcal{R}_{jk}}{M_{\Psi_k} \mathcal{I}^d_k} \right).
\label{ampoct}
\ee
One can immediately see from this expression that for degenerate
fermion octets, i.e., for $M_{\Psi_k}=M_{\Psi}$ the elements of $\mathcal{R}$ drop out of
this expression and the amplitude is proportional to $M_{ee} = \sum m_i
U^2_{ei}$. Therefore, if the fermion octets are degenerate, both
the direct as well as the indirect 
contributions to \obb~are directly proportional to the effective
mass $M_{ee}$. As a result, if due to any cancellation the contribution from light neutrino 
exchange goes to zero, the corresponding contribution from the octet exchange 
also vanishes identically. However barring these cancellations due the Majorana 
phases, we will see that the relative importance of the two contributions may widely vary depending on the model parameters.

Let us point out that in some sense \obb\ and the generation of
neutrino masses can be decoupled in this model. 
It can be seen from Eqs.~(\ref{eq:lagrangian-nus}) and
(\ref{eq:lagrangian-quarks}) that two independent sources of lepton
number violation exist. The first term of Eq.~(\ref{eq:lagrangian-nus}) may be used to
assign lepton number to the color octet particles, and the
last two terms then break it. If $\lambda_{\Phi H}$ 
is zero, we can assign lepton number either to $\Phi$ or to
$\Psi_i$. In the first case (assigning
lepton number to $\Phi$) the quark couplings to $\Phi$ in 
Eq.~(\ref{eq:lagrangian-quarks}) provide LNV; in the second case (assigning
lepton number to $\Psi_i$), the colored fermion mass term gives
LNV.  If now $\lambda_{\Phi H} = 0$, the one-loop neutrino masses in
Eq.~(\ref{eq:numass}) vanish\footnote{Since lepton number is not
conserved, higher order diagrams will lead to very small neutrino
masses.}, but still there are non-vanishing 
contributions to \obb.  

\begin{figure}
 \centering
\includegraphics[scale=1]{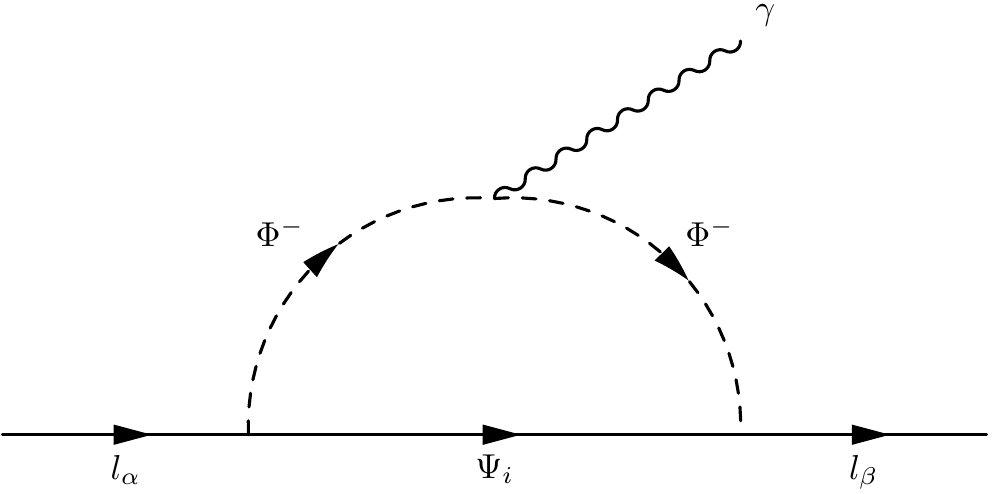}
\caption{Lepton flavor violating process $l_\alpha \to l_\beta 
\gamma$ in the colored seesaw scenario.}\label{litoljgamma}
\end{figure}

Besides neutrinoless double beta decay, the color octets can actively
participate in different lepton flavor violating processes, like $\mu
\to e\gamma$, $\tau \to e\gamma$ and $\tau \to \mu \gamma$, as shown
in Fig.~\ref{litoljgamma}. The branching ratio corresponding to
$l_\alpha \to l_\beta \gamma$ is
\cite{FileviezPerez:2009ud,FileviezPerez:2010ch, Liao:2009fm}
\be
\mathrm{Br}(l_\alpha \to l_\beta \gamma)= \frac{3 \alpha_\mathrm{em}}{ 4\pi G^2_F M^4_{\Phi}}\left | \sum_{i}Y^{\beta i}_{\nu}
(Y^{\alpha i}_{\nu})^* \mathcal{F}(x_i)\right|^2 ,
\ee
where the function $\mathcal{F}(x_i)$ has the following form, 
\be
\mathcal{F}(x_i)=\frac{1-6x_i+3 x_i^2+ 2x_i^3-6 x^2_i \ln(x_i)}{12(x_i-1)^4}\,,
\ee
with $ x_i = M_{\Psi_i}^2/M_\Phi^2$ 
and  $\alpha_\mathrm{em}$ being the fine structure constant.  
As for the case of the \obb\ amplitude, one can write the branching ratio of
$\mu \to e\gamma$ in terms of the neutrino parameters and the matrix $\mathcal{R}$ as 
\be \label{eq:lfv}
\mathrm{Br}(\mu \to e\gamma) =\frac{3 \alpha_\mathrm{em}}{ 4\pi G^2_F M^4_{\Phi}} \frac{ (16 \pi^2)^2}{\lambda^2_{\Phi H} v^4} \left| 
\sum_k \frac{\mathcal{F}(x_k)}{\mathcal{I}_k}
 \sum_{i,j} U_{ei} U^*_{\mu j} \mathcal{R}_{ik } \mathcal{R}^*_{jk } 
\sqrt{m_{i} m_j} \right|^2 \,.
\ee
For the simple choice of $\mathcal{R}_{ij}=\delta_{ij}$, the above expression reduces to
\be
\mathrm{Br}(\mu \to e\gamma) = \frac{3 \alpha_\mathrm{em}}{ 4\pi G^2_F M^4_{\Phi}} \frac{ (16 \pi^2)^2}{\lambda^2_{\Phi H} v^4}  \left| 
\sum_i \frac{\mathcal{F}(x_i)}{\mathcal{I}_i}
  U_{ei} U^*_{\mu i}  m_i \right|^2 .
\ee
In fact, one can check that the branching ratio is independent of $\mathcal{R}$
if the fermion octets are degenerate and $\mathcal{R}$ is real. Since 
$\mathcal{R}$ embodies our lack of knowledge of the seesaw-scale physics 
which cannot be determined from low-scale data, we can immediately see 
that for such a class of model parameters the theory is fully determined 
leading to unambiguous prediction for lepton flavor violating decays.
Currently, the best limit on lepton flavor violation 
comes from the MEG collaboration, which gives \cite{MEG} 
\begin{equation}
 \mathrm{Br}(\mu \rightarrow e\gamma) \leq 2.4 \times 10^{-12} 
\end{equation}
at 90\% C.L. In what follows we will discuss some simple examples for
the interplay of neutrino mixing, neutrinoless double beta decay and
lepton flavor violation. While this is not a complete analysis, some
very interesting and general features arise.

\section{ Neutrinoless Double Beta Decay and Lepton Flavor Violation with Two Color Octet Fermions \label{twodeg}}

In this section, we discuss neutrinoless double beta decay and 
lepton flavor violating processes, considering
the minimal case with two degenerate color octet fermions. 
Thus the matrix ${\cal R}$ depends on only one complex parameter and 
can be taken as 
\be
\mathcal{R} (\rm {for \,\,NH})=\pmatrix{ 0 & 0 \cr \sqrt{1-\omega^2} & -\omega \cr \omega & \sqrt{1-\omega^2}}\mbox{ and }
\mathcal{R} (\rm{for \,\,IH})=\pmatrix {\sqrt{1-\omega^2} & -\omega \cr \omega & \sqrt{1-\omega^2} \cr 0 & 0} .
\ee
In our analysis we consider a real $\omega$ and hence, $-1 \le \omega \le +1$. Note that
since we need one heavy fermion for each light neutrino, 
this results in the lightest neutrino mass being zero. 
We first discuss the normal hierarchy and then the inverted hierarchy scenario.

 \subsection{Normal Hierarchy}

For the normal hierarchy scenario and two color octet fermions with degenerate masses, i.e., 
$M_{\Psi_i}=M_{\Psi}$ (and hence ${\mathcal{I}_i} = {\mathcal{I}}$) for all $i$, 
the Yukawas can be expressed as 
\be
Y_{\nu}=\sqrt{\frac{16 \pi^2}{\lambda_{\Phi H} }}\frac{1}{v} U \,
\mathrm{diag}(0, \sqrt{m_2}, \sqrt{m_3}) \, \mathcal{R} \,
\mathrm{diag}(\sqrt{\mathcal{I}^{-1}}, \sqrt{\mathcal{I}^{-1}})\, .
\ee
Expressed  in terms of the elements of the PMNS mixing matrix $U$,
the light neutrino mass eigenvalues $m_i$ and the quartic coupling $\lambda_{\Phi H}$, the corresponding half-life for 
\obb~can be
written as
\be
\frac{1}{T_{1/2}}= \mathcal{K}_{0 \nu}  \left( \frac{1}{P^2 }+ \frac{
\tilde{y}^2_{11}}{M_{\Psi} M^4_{\Phi} G^2_F}\frac{ 16 \pi^2}{v^2
\lambda_{\Phi H} \mathcal{I}} \right)^2 \left| m_2 U^2_{e2}+ m_3
U^2_{e3} \right |^2 .
\ee
As discussed in the previous section, since we have considered
degenerate fermions, the expression for the half-life of neutrinoless
double beta decay is independent of the parameter $\omega$, although
the  Yukawa couplings  depend strongly on it.  
\begin{figure}[t]
\begin{center}
\includegraphics[width=0.45\textwidth]{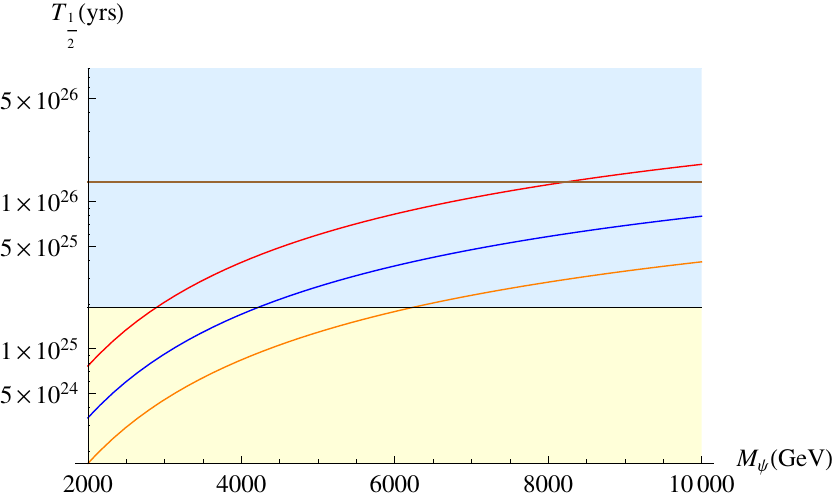}
\includegraphics[width=0.45\textwidth]{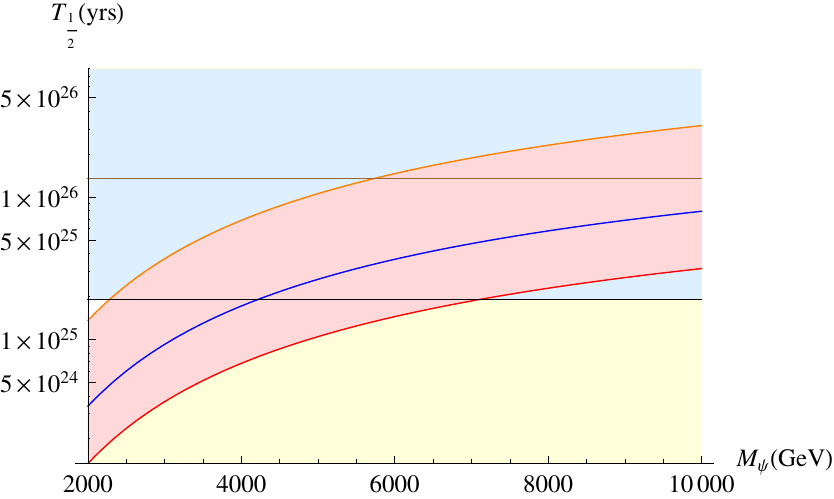}
\end{center}
\caption{Half-life $T_{1/2}$ of $^{76}$Ge  against  the mass of the color octet 
fermion $M_{\Psi}$. 
The mass of scalar octet has been fixed as $M_{\Phi}=1.96$ TeV. In the
left plot, the red, blue and orange lines correspond 
 to $\lambda_{\Phi H} =0.15 \times 10^{-8}$, $\lambda_{\Phi H} = 0.1\times 10^{-8}$, and $\lambda_{\Phi H} =0.7 \times 10^{-9}$, respectively. 
The yellow region is  experimentally excluded. The black line
corresponds to the Heidelberg--Moscow limit. The 
brown line represents the light neutrino contribution to
\onbb~divided by  $50^2$, for
the neutrino oscillation parameters mentioned in the text. In the
right plot the red, blue and orange lines correspond
 to the three different nuclear matrix elements $\mathcal{M}_{\Phi \Psi}=600.38$, $\mathcal{M}_{\Phi \Psi}=377.59$, and $\mathcal{M}_{\Phi \Psi}=188.79$, 
respectively. The coupling has been set to $\lambda_{\Phi H} = 10^{-9}$. 
Both figures are for normal hierarchy.
}
\label{figg}
\end{figure}
We also stress again that for the case of degenerate octet fermions
both the light neutrino contribution and the color octet
contributions share the same proportionality factor  $|M_{ee}|=\left|
m_2 U^2_{e2}+ m_3 U^2_{e3} \right |$. Hence, if  due to any
cancellation between neutrino parameters the  light neutrino
contribution becomes zero, the same cancellation will set the color
octet contribution to zero, too. However, for $M_{ee} \neq 0$ 
the direct contribution from the color octets can be taken 
independently of the neutrino mass scale. 
This is because there is an additional parameter $\lambda_{\Phi H}$ 
in addition to the Yukawa coupling involved, and hence one can suitably 
adjust the two to get small enough neutrino masses required for normal 
hierarchy and yet get a very large contribution to \obb. 
In particular, by choosing the coupling $\lambda_{\Phi H}$ to smaller values,
the color octet contribution can dominate over the light neutrino contribution.
We have shown the relative comparison between the light neutrino and
color octet contributions in Fig.~\ref{figg}, where we have plotted
the variation of the half-life $T_{1/2}$  vs.\ the mass of the color
octet fermion $M_{\Psi}$.  The details of the figure are as follows: 
\begin{itemize}
\item
We used the neutrino oscillation parameters 
$\Delta m^2_{21}= 7.5 \times 10^{-5}\, \rm{eV}^2$, $\Delta m^2_{31}= 2.3 \times 10^{-3} \, \rm{eV}^2$, $\theta_{12}=33^{\circ}$, $\theta_{23}
=42^{\circ}$ and $\theta_{13}=8^{\circ}$, see
Ref.~\cite{Schwetz:2008er} for a recent global fit. For this choice of parameters, the effective mass  is $M_{ee}= 0.003$ eV. The elements $U_{e2}$ and $U_{e3}$ have been considered
real. 
\item
The mass of the scalar octet has been fixed as $M_{\Phi}=1.96$ TeV, so
that it satisfies the present bound on the mass of color octet scalars
\cite{LHC}. However, in \cite{LHC} the bound has been derived assuming negligible
coupling with quarks, i.e., production in the gluon fusion channel. In our case, the 
color octet scalar has interaction with the quarks, hence 
the mass bound may be weakened.
\item
The black line represents the Heidelberg--Moscow limit \cite{epj} for
neutrinoless double beta decay and corresponds to $T_{1/2}=1.9 \times
10^{25}$ yr. The yellow region is thus excluded, while the blue
region is the allowed one.
\item
The brown line represents the standard light neutrino contribution to
\obb~scaled by a factor $50^2$, i.e., $ {\mathcal{K}^{-1}_{0 \nu} }\left(\frac{M_{ee}}{P^2}\right)^{-2} \times \frac{1}{50^2}$, where
we have taken $P^2 \sim (251.49)^2 \, 
\rm{MeV}^2$. 
The factor $\mathcal{K}_{0\nu}$ depends on the nuclear matrix 
element and can be obtained using $\mathcal{K}_{0\nu}= G_{0
\nu}(\mathcal{M}_{\Phi \Psi} m_p)^2$. 
We have taken $\tilde{y}^2_{11} = 1$ and adopted the nuclear matrix
elements from  \cite{pion-ex, hirsch, allanach}, where $\mathcal{M}_{\nu}=2.8$, and following Eq.~(A.8) in \cite{allanach},  $\mathcal{M}_{\Phi \Psi}=377.59$, if the pion exchange and usual one and two nucleon mode are considered. 
For this value of ${M}_{\Phi \Psi}$, $\mathcal{K}_{0 \nu}=9.95
\times 10^{-10} \, \rm{ yr}^{-1} \, \rm{GeV}^2$ and  the standard contribution is 
  ${\mathcal{K}^{-1}_{0 \nu} }\left(\frac{M_{ee}}{P^2}\right)^{-2}=3.38\times 10^{29} \, \rm{yr}$.
 These \obb-matrix element values are
the ones which have been evaluated for the case of short range
$R$-parity violating SUSY diagrams, which is closest to the setup
considered by us. In particular, the operator structure 
which corresponds to $\lambda^{\prime}_{111}$ diagrams is similar to our case.
Hence the information regarding nuclear matrix element for $R$-parity 
violation can be utilized in our scenario.

\item
The red, blue and orange  lines correspond to the values $\lambda_{\Phi H} = 0.15 \times 10^{-8}$, $\lambda_{\Phi H} = 0.1\times 10^{-8}$ and $\lambda_{\Phi H} = 0.7 \times 10^{-9}$, respectively. The half-life 
decreases with $\lambda_{\Phi H} $.
\item In Fig.~\ref{figg}, we have also shown the variation of $T_{1/2}$ 
with the mass of the color octet fermion, considering
the variation of nuclear matrix element $\mathcal{M}_{\Phi \Psi}$ in the interval $188.79-600.38$.
\end{itemize}
The figure clearly shows that the direct octet contribution can dominate the  
contribution from light neutrino exchange easily, which, we reiterate, has been shown by 
dividing  by a factor of $(50)^2$. It is also clear that the direct octet contribution can 
easily saturate the current limit on the half-life of \obb, even for normal 
hierarchy. Note that the prediction for \obb\, with light Majorana 
neutrino exchange for normal hierarchy is very low compared to 
that for inverted hierarchy, and much below the reach of the next generation 
\obb\, experiments. However, since in the colored seesaw model one can have 
very large \obb\,  even for normal hierarchy, \obb\, can no longer be used distinguish between 
the neutrino mass hierarchies.

\begin{figure}[t]
\begin{center}
\includegraphics[width=0.5\textwidth]{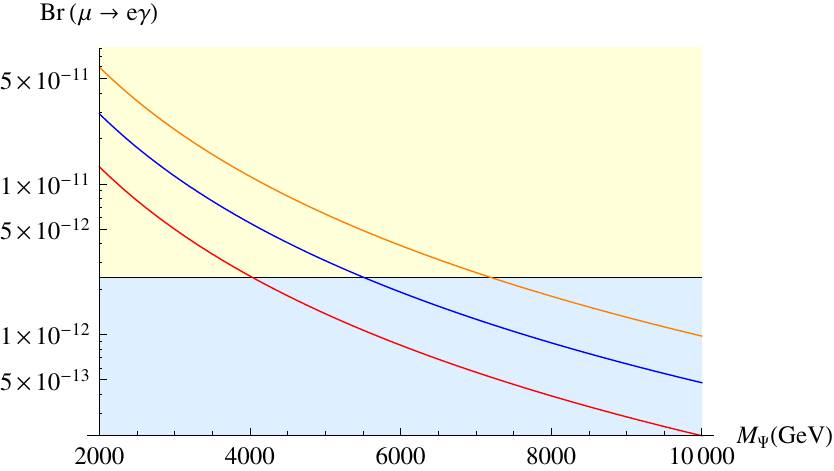}
\end{center}
\caption{Branching ratio of $\mu \rightarrow e \gamma$ as a function
of the octet fermion mass $M_\Psi$ for normal hierarchy. The mass of
scalar octet has been fixed as $M_{\Phi}=1.96$ TeV.  The red, blue and
orange lines correspond  to  three different values $\lambda_{\Phi H} =
0.15 \times 10^{-8}$, $\lambda_{\Phi H} = 0.1\times 10^{-8}$, and
$\lambda_{\Phi H} = 0.7 \times 10^{-9}$. The yellow region is
experimentally excluded. The black line corresponds to the present
bound from the MEG experiment. }
\label{figlfv1}
\end{figure}

Having discussed lepton number violation, we turn to lepton flavor
violation now.  
We are interested in exploring if the color-octet fermions can 
simultaneously produce a $0\nu \beta\beta$ rate large enough to saturate the 
current limit, and at the same time saturate the  experimental bound for $\mu
\to e\gamma$. The branching ratio for this process has 
been given in Eq.~(\ref{eq:lfv}). We stick to real $\omega$ and degenerate
fermions and hence the branching ratio
becomes independent of the parameter $\omega$. It has the following
form: 
\be
\mathrm{Br}(\mu \to e\gamma)= \frac{3 \alpha_\mathrm{em}}{ 4\pi G^2_F
M^4_{\Phi}} \frac{ (16 \pi^2)^2}{\lambda^2_{\Phi H} v^4} \left(\frac{{
\mathcal{F}(x)}}{\mathcal{I}}\right)^2 \left| m_2 U_{e2} U^*_{\mu 2}+
m_3 U_{e3} U^*_{\mu 3} \right|^2 . 
\ee
Like in neutrinoless double beta decay, the color octet fermions can
also  give a significant contribution to this process. The result has been 
shown in Fig.~\ref{figlfv1}, where we have given the variation 
of the branching ratio against the mass of the color octet fermions,
considering three different values of $\lambda_{\Phi H}$. The summary
of the figure is as follows: 
\begin{itemize}
\item
The neutrino oscillation parameters as well as  the mass of scalar octet   have been set to the previously mentioned values, already used in the $0\nu \beta\beta$ study. For $\theta_{12} =33^{\circ}$, $\theta_{23}=42^{\circ}$, $\theta_{13}=8^{\circ}$, 
$\Delta m^2_{21}= 7.5 \times 10^{-5} \, \rm{eV}^2$ and $\Delta m^2_{31}= 2.3 \times 10^{-3}\, \rm{eV}^2$, we have
$(m_2 U_{e2} U_{\mu 2}+ m_3 U_{e3} U_{\mu 3}) = 7.1 \times
10^{-12} \, \rm{GeV}$. The elements of the mixing matrix have been
considered real.
\item
The black line corresponds to the present bound coming from  $\mu \to
e \gamma$ searches at MEG experiment, i.e., $ {\rm Br} (\mu
\to e \gamma) = 2.4 \times 10^{-12}$.
\item
The red, blue and orange lines correspond to the values $\lambda_{\Phi H} = 0.15 \times 10^{-8}$, $\lambda_{\Phi H} = 0.1\times 10^{-8}$,
and $\lambda_{\Phi H} = 0.7 \times 10^{-9}$, respectively. 
\end{itemize}
It is clear from the figure that one can rather easily saturate the
current lepton flavor violation limits. Note that in principle the
expression $\left| m_2 U_{e2} U^*_{\mu 2}+ m_3 U_{e3} U^*_{\mu 3}
\right|$, and thus the branching ratio for $\mu \to e \gamma$, can
vanish if the mixing parameters and CP phase conspire. This would also
be possible when we considered complex $\mathcal{R}$.

\begin{figure}[t]
\includegraphics[width=0.495\textwidth]{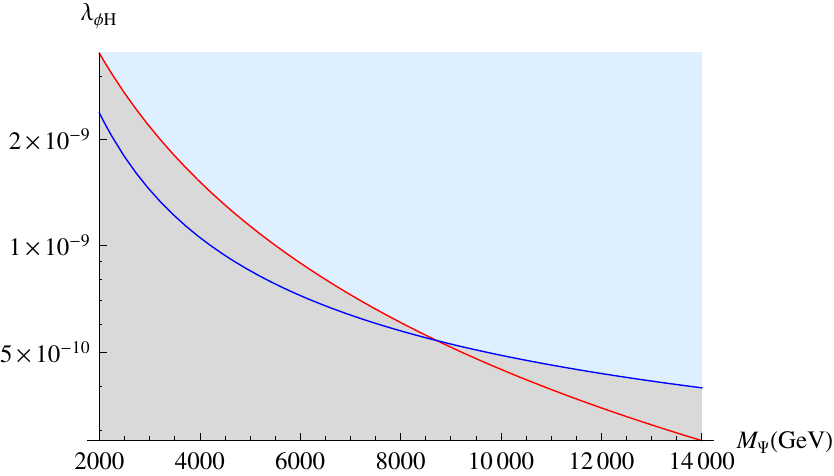}
\includegraphics[width=0.495\textwidth]{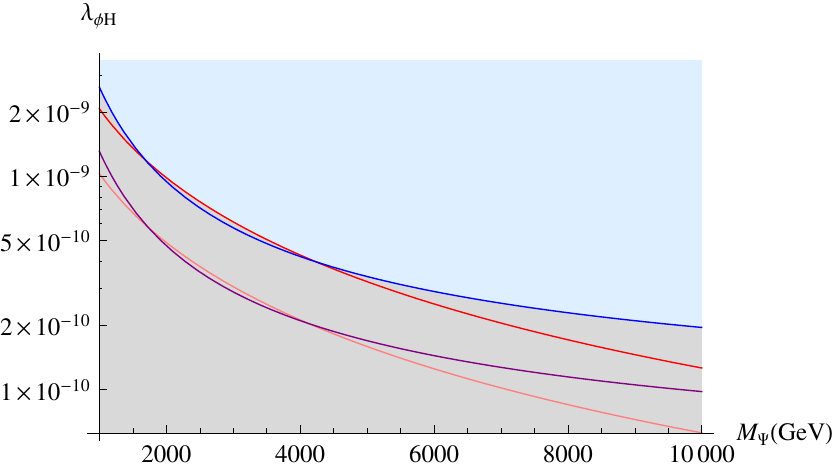}
\caption{The coupling $\lambda_{\Phi H}$ against the mass of the color octet fermions $M_{\Psi}$.
The mass of scalar octet has been fixed as $M_{\Phi}=1.96$ TeV.  The
gray band is experimentally excluded by $\mu \to e\gamma$ and $0\nu
\beta\beta$. Left panel: the red line represents the present limit
obtained by the MEG experiment for $\mu \to e\gamma$. The blue
line represents the Heidelberg--Moscow bound. The factor
$\tilde{y}^2_{11}$ has been considered to be $1$.  Right panel:
The blue and purple lines correspond to the Heidelberg--Moscow bound, while the $\tilde{y}^2_{11}$ factor has been considered as
0.4 and 0.2 respectively. The red and pink lines correspond to the MEG
limit, while we have considered a suppression factor 0.08
and 0.0197 from the factor $ \left |m_2 U_{e2} U^*_{\mu 2 }+ m_3
U_{e3} U^*_{\mu 3} \right |^2$. The  figure has been generated
considering normal hierarchy. 
}
\label{contour-a}
\end{figure}

If the color octet contribution saturates both the Heidelberg--Moscow
bound, as well as the present bound on $\mu \to e\gamma$ decay coming
from MEG experiment, then the mass of the scalar and fermionic
octets,  and the coupling $\lambda_{\Phi H}$ should satisfy
simultaneously the following two equations: 
\bead \displaystyle
\lambda_{\Phi H} &= &  \displaystyle 1.63\times 10^{20} \,\frac{\sqrt{\mathcal{K}_{0 \nu}}}{ M_{\Psi} M^4_{\Phi}}\,  \tilde{y}^2_{11} \frac{\left| m_2 U^2_{e2}+ 
m_3 U^2_{e3} \right|}{ \mathcal{I}}  \, , \\
\lambda_{\Phi H} & =&  \displaystyle 1.18 \times 10^{7} \, \frac{
\mathcal{F}(x)}{\mathcal{I}} \, \frac{\left| m_2 U_{e2} U^*_{\mu 2}+
m_3 U_{e3} U^*_{\mu 3} \right|}{M^2_{\Phi}} \, . 
\eea
We have represented the above two conditions in Fig.~\ref{contour-a}, where we have shown 
the variation of $\lambda_{\Phi H}$ against the mass of 
the color octet fermions $M_{\Psi}$. As before, the mass of the scalar octet and 
 $\mathcal{K}_{0\nu}$ have been set to be the same value as for Fig.~\ref{figg}. For simplicity,
we have considered
the elements of the PMNS mixing matrices to be real. In Fig.~\ref{contour-a}, the gray region is 
excluded both from $0\nu \beta\beta$ and $\mu \to e\gamma$. The red and blue lines in the left 
panel  correspond to $\mathrm{Br} (\mu \to e\gamma)=2.4 \times 10^{-12}$, and the $0\nu \beta\beta$ 
saturating bound $T_{1/2}=1.9 \times 10^{25}\, \rm{yr}$,
where the Yukawas between scalar octets and quarks have been considered $\mathcal{O}(1)$. 
The same exercise has been repeated for the right panel of  Fig.~\ref{contour-a} with a different $\tilde{y}^2_{11}$ factor. Note that with the decrease of $\tilde{y}_{11}^2$, one will require an
additional suppression in $ \left| m_2 U_{e2} U^*_{\mu 2}+ m_3
U_{e3} U^*_{\mu 3} \right|$ in order to simultaneously saturate the MEG and
Heidelberg--Moscow limit. This can come from cancellation between the phases. 
Also, interestingly  Fig.~\ref{contour-a} indicates that if only lepton flavor 
violation is saturating, the color octet fermions can be within the reach of 
LHC. However, inclusion of a saturating $0\nu \beta \beta$ demands the color-octet
fermion mass to be higher. 

\subsection{Inverted Hierarchy}

\begin{figure}[t]
\includegraphics[width=0.495\textwidth]{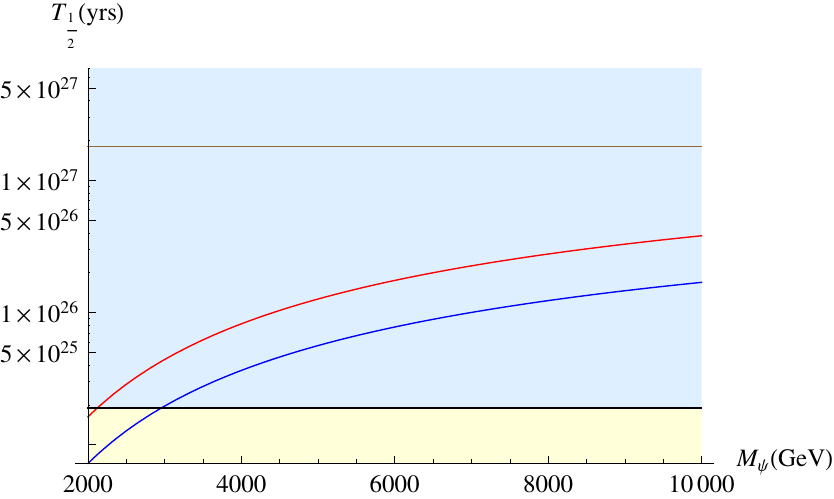}
\includegraphics[width=0.495\textwidth]{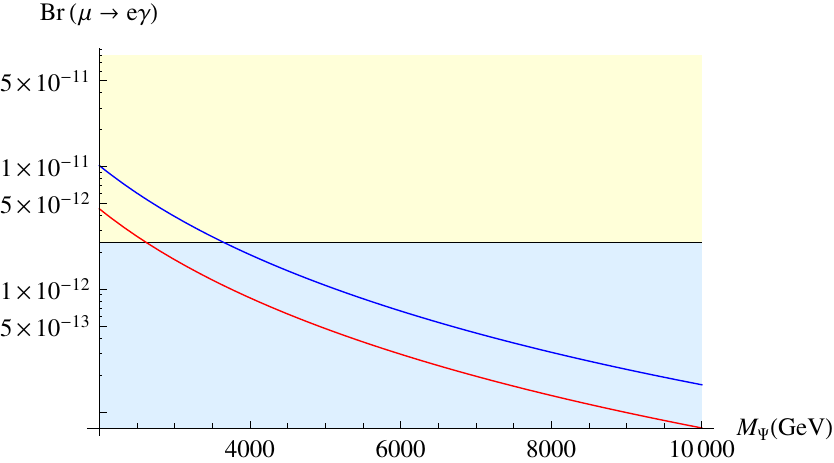}
\caption{
Half-life of $0\nu\beta\beta$ in $^{76}$Ge and the branching ratio for
$\mu \rightarrow e \gamma$ as a function of the octet fermion mass
$M_\Psi$ for inverted hierarchy. The mass of scalar octet has been
fixed as $M_{\Phi}=1.96$ TeV.   The red and  blue  lines correspond to $\lambda_{\Phi H} = 0.15 \times
10^{-8}$ and $\lambda_{\Phi H} = 0.1\times 10^{-8}$, respectively. In the left panel the
light neutrino  contribution has been shown in gray, without any
scaling factor. While for the left plot $\tilde{y}^2_{11}=0.05$, the
right panel is generated without any additional suppression factor. 
The blue areas are allowed, the yellow ones forbidden.
}
\label{figlfv3}
\end{figure}

\begin{figure}[t]
\begin{center}
\includegraphics[width=0.5\textwidth]{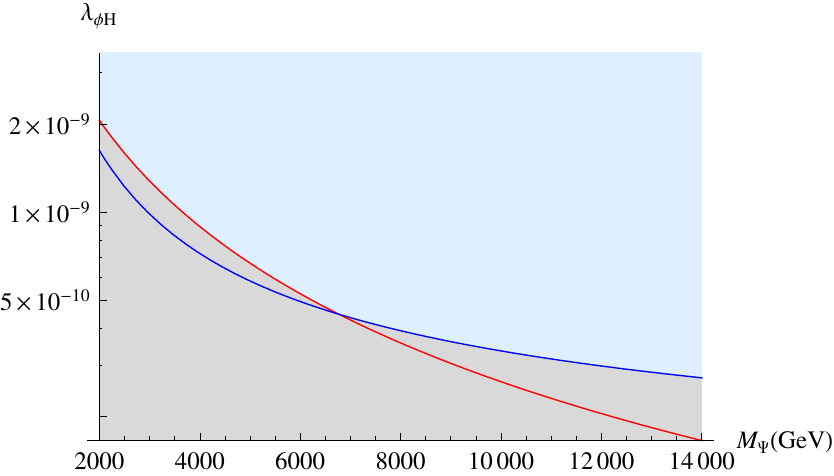}
\end{center}
\caption{
The coupling $\lambda_{\Phi H}$ against the mass of the color octet fermions $M_{\Psi}$ for inverted hierarchy.  The
gray band is experimentally excluded by $\mu \to e\gamma$ and $0\nu
\beta\beta$. The mass of scalar octet has been fixed as $M_{\Phi}=1.96$ TeV.  The red and  blue lines represent
the saturating contribution in MEG  experiment and Heidelberg--Moscow 
bound. The factor $\tilde{y}^2_{11}$ has been set to 0.05. }
\label{figlfv}
\end{figure}

We discuss briefly the relative comparison of neutrinoless double
beta decay and $\mu \to e \gamma$ when the light neutrino
states follow the inverted hierarchy. The half-life for \obb~can then be expressed as
\be
\frac{1}{T_{1/2}}=\mathcal{K}_{0 \nu} \left( \frac{1}{P^2 }+ \frac{ \tilde{y}^2_{11}}{M_{\Psi} M^4_{\Phi} G^2_F}\frac{ 16 \pi^2}{v^2 \lambda_{\Phi H} \mathcal{I}} \right)^2 \left| m_1 U^2_{e1}+ m_2 U^2_{e2} \right |^2 ,
\ee
and the branching ratio for $\mu \to e\gamma$ is given by
\be
\mathrm{Br}(\mu \to e\gamma)= \frac{3 \alpha_\mathrm{em}}{ 4\pi G^2_F M^4_{\Phi}} \frac{ (16 \pi^2)^2}{\lambda^2_{\Phi H} v^4} \left(\frac{{ \mathcal{F}(x)}}{\mathcal{I}}\right)^2 \left| m_1 U_{e1} U^*_{\mu 1}+ m_2 U_{e2} U^*_{\mu 2} \right|^2 \, .
\ee
Since, in general even for light neutrino states, the contribution for
inverted hierarchy  is larger than the contribution for normal
hierarchy, the same  feature holds for the colored seesaw 
scenario with degenerate fermions. One can obtain a significantly
large neutrinoless double beta decay contribution with relatively
small values of quark Yukawas, and/or larger color octet scalar and
fermion masses. For completeness, we show two figures to support this 
feature. See Figs.~\ref{figlfv3} and \ref{figlfv} for more
details. All parameters except the ones explicitly mentioned in the
captions are the same as for the previous figures.

We would like to end this section by discussing the particular
relations which the different lepton flavor violating processes share
among themselves in the special case considered here.  The
ratio of the branching ratios for different lepton flavor violating processes are 
\be\label{eq28}
\frac{{\rm Br}(\tau \to e\gamma)}{{\rm Br}(\mu \to e\gamma)} =\frac{ 
\left( U_{e2} U^*_{\tau 2} m_2 + U_{e 3}U^*_{\tau 3} m_3 \right)}
{\left( U_{e2} U^*_{\mu 2} m_2 + U_{e 3}U^*_{\mu 3} m_3 \right)} ~,~~
\frac{{\rm Br}(\tau \to \mu \gamma)}{{\rm Br}(\mu \to e\gamma)}=\frac{ \left( U_{\mu 2} U^*_{\tau 2} m_2 + U_{\mu 3}U^*_{\tau 3} m_3 \right)}{\left( U_{e2} U^*_{\mu 2} m_2 + U_{e 3}U^*_{\mu 3} m_3 \right)}
\ee
for normal hierarchy, and 
\be\label{eq29}
\frac{{\rm Br}(\tau \to e\gamma)}{{\rm Br}(\mu \to e\gamma)}=\frac{ 
\left( U_{e1} U^*_{\tau 1} m_1 + U_{e 2}U^*_{\tau 2} m_2 \right)}
{\left( U_{e1} U^*_{\mu 1} m_1 + U_{e 2}U^*_{\mu 2} m_2 \right)} ~,~~
\frac{{\rm Br}(\tau \to \mu \gamma)}{{\rm Br}(\mu \to e\gamma)}=\frac{ 
\left( U_{\mu 1} U^*_{\tau 1} m_1 + U_{\mu 2}U^*_{\tau 2} m_2 \right)}{\left( U_{e1} U^*_{\mu 1} m_1 + U_{e 2}U^*_{\mu 2} m_2 \right)}
\ee
for inverted hierarchy. 
Note that, since for the two octet fermion case with real ${\cal R}$ there is no additional 
parameter in the theory apart from the ones measured in low energy experiments, one can 
get an exact prediction for these ratios of branching ratios in terms of oscillation 
parameters. Hence, experimental measurement of some or all of them can be used to 
check if the lepton flavor violation is solely due to the neutrino mass generation mechanism 
or not. 
The present bounds are $\mathrm{Br}(\tau \to \mu \gamma) \leq 4.5 \times 
10^{-8}$ and $\mathrm{Br}(\tau \to e\gamma) \leq 1.2 \times 10^{-7}$, respectively \cite{belle}. 
 Disregarding the possibility of cancellations in the numerator or denominator,
the above expressions (\ref{eq28}) and (\ref{eq29}) are expected to be of order one.  Since experimentally the limits obey ${\rm Br}(\mu \to e\gamma) \ll {\rm
Br}(\tau \to \mu\gamma)$ as well as ${\rm Br}(\mu \to e\gamma) \ll {\rm Br}(\tau \to e\gamma)$, the
limits on $\tau \to \mu \gamma$ and $\tau \to e\gamma$ are therefore
automatically obeyed if the limit on $\mu \to e\gamma$ is obeyed.

\section{Neutrinoless Double Beta Decay  with Three Color Octet Fermions \label{threegen}}
After discussing lepton number and lepton flavor violating processes
for two degenerate color octet fermionic states in the last section,
we now turn to the three generation scenario, dropping in addition the
assumption of degenerate octet fermions. 
It is evident from Eq.~(\ref{ampoct}) and the discussion following it
that in this
case the color octet contribution and the light neutrino contribution
of \obb~do not share the same proportionality factor $M_{ee}$ anymore. This particular 
feature brings up the possibility that, even if the light neutrino
contribution becomes zero due to cancellation between the terms with
$m_1$, $m_2$ and $m_3$, the color octet contribution can be non-zero,
and even significantly large. We explicitly show this specific
feature for one case. 

In general, the color octet contribution will have a significant dependence 
on the phases of the matrix $\mathcal{R}$. We do not address this
issue in the present work, as we will encounter extreme cases even
when $\mathcal{R}_{ij}=\delta_{ij}$, in which case the particle
physics amplitude of \obb~is given by
\be
\mathcal{A} \simeq \frac{16 \pi^2}{\lambda_{\Phi H} v^2} \frac{y^2_{11}}{M^4_{\Phi}} \left( \sum_i \frac{ m_i U^2_{ei}  }
{M_{\Psi_i} \mathcal{I}_i} \right) . 
\ee
We will stick to this simple case throughout the rest of this section, as very interesting features arise already at this stage.

In what regards neutrino mixing, we try to keep things as simple as
possible, and study a somewhat minimal deviation from tri-bimaximal
mixing. Denoting $\sin \theta_{13}=\lambda$, the PMNS mixing matrix is now
\begin{equation}
U_\mathrm{PMNS} \simeq \pmatrix{
                          -\frac{2}{\sqrt{6}} & \frac{1}{\sqrt{3}} & \lambda e^{-i\delta} \cr
                          \frac{1}{\sqrt{6}}-\frac{\lambda}{\sqrt{3}}e^{i\delta} &  \frac{1}{\sqrt{3}}+\frac{\lambda}{\sqrt{6}}e^{i\delta} & -\frac{1}{\sqrt{2}} \cr
                           \frac{1}{\sqrt{6}} +\frac{\lambda}{\sqrt{6}}e^{i\delta} & \frac{1}{\sqrt{3}}-\frac{\lambda}{\sqrt{6}}e^{i\delta}& \frac{1}{\sqrt{2}}
                         }
\mathrm{diag}(1, e^{i\alpha},e^{i (\beta + \delta)}) \, .
\end{equation}
 Note that we allow for a complex PMNS matrix in what follows. In this case, the amplitude for the light neutrino contribution to $0\nu\beta\beta$ is
\begin{equation}
 \mathcal{A}_l \simeq \frac{G_F^2}{\langle p^2\rangle}\left( \frac{2m_1}{3} + \frac{m_2}{3} e^{i 2 \alpha} + 
m_3 \lambda^2 e^{i 2 \beta} \right)  .
\end{equation}
For the amplitude of $0\nu\beta\beta$ mediated by the color octet fermions and scalars, we have 
\begin{equation}\label{amp3a}
 \mathcal{A} \simeq \frac{y^2_{11}}{M^4_{\Phi}} \frac{16 \pi^2}{\lambda_{\Phi H} v^2} 
\left( \frac{2 m_1}{3 \mathcal{I}_1 M_{\Psi_1}} +\frac{m_2 e^{i 2 \alpha}}{3 \mathcal{I}_2 M_{\Psi_2}} + 
\frac{m_3 \lambda^2 e^{i 2 \beta}}{\mathcal{I}_3 M_{\Psi_3}}\right)  .
\end{equation} 
The branching ratio for the process $\mu \rightarrow e \gamma$ is given by
\be
 \mathrm{Br}(\mu \rightarrow e \gamma) \propto   \left| ( 2 e^{i\delta} \lambda -\sqrt{2} ) \frac{m_1 \mathcal{F}(x_1)}{\mathcal{I}_1} \right. 
\left. + (\sqrt{2} e^{i 2 \alpha} + \lambda e^{i 2 \alpha + i\delta} ) \frac{m_2 \mathcal{F}(x_2)}{\mathcal{I}_2} -
 3 e^{i 2 \beta+i \delta} \lambda \frac{m_3 \mathcal{F}(x_3)}{\mathcal{I}_3} \right|^2 \, ,
\ee
where the proportionality factor is $\frac{2}{3} \frac{16 \pi^3}{\lambda_{\Phi H}^2 v^4} \frac{\alpha_\mathrm{em}}
{G_F^2 M^4_{\Phi}}$.

\begin{figure}[t]
\centering
\begin{minipage}{0.4\linewidth}
\centering
\includegraphics[width=\textwidth]{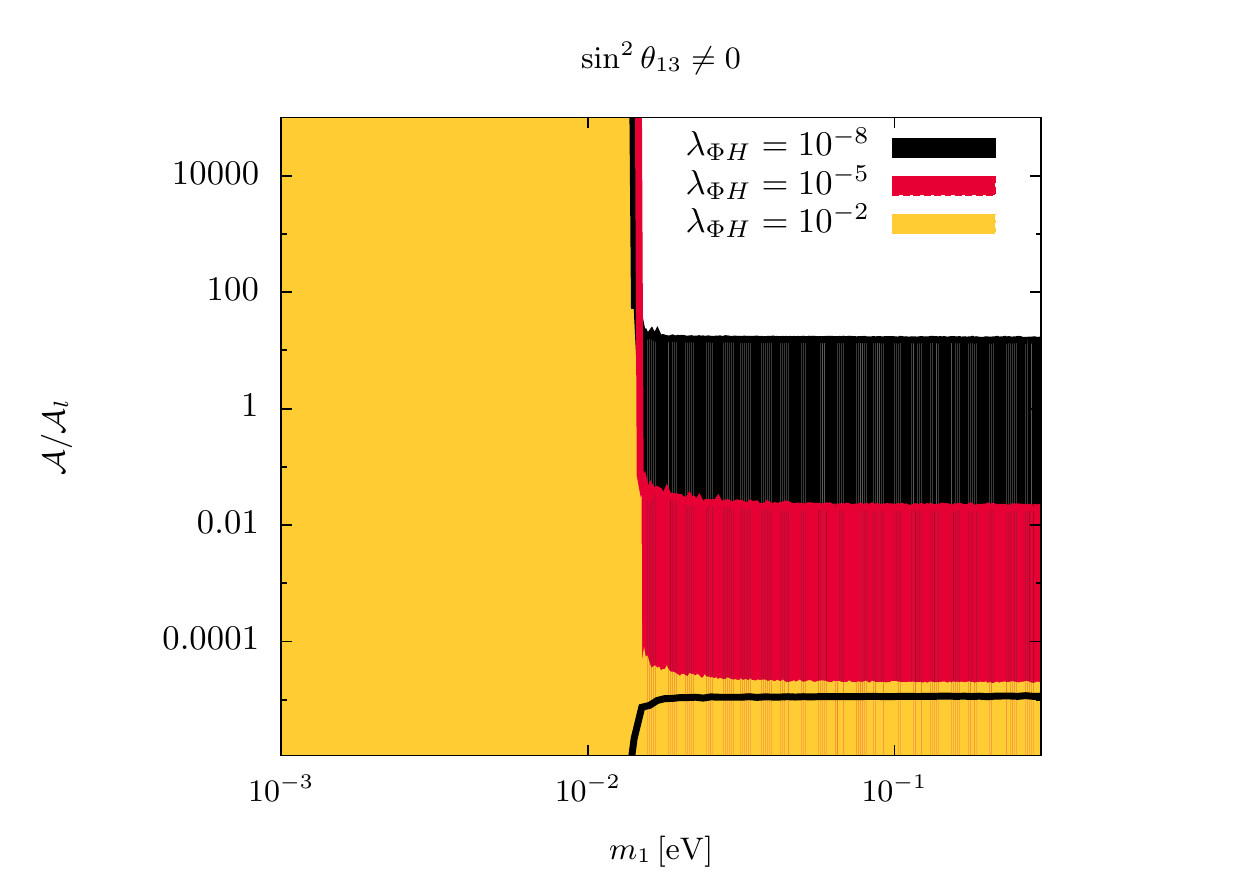}
\end{minipage}
\hspace{.1\linewidth}
\begin{minipage}{0.4\linewidth}
\centering
\includegraphics[width=\textwidth]{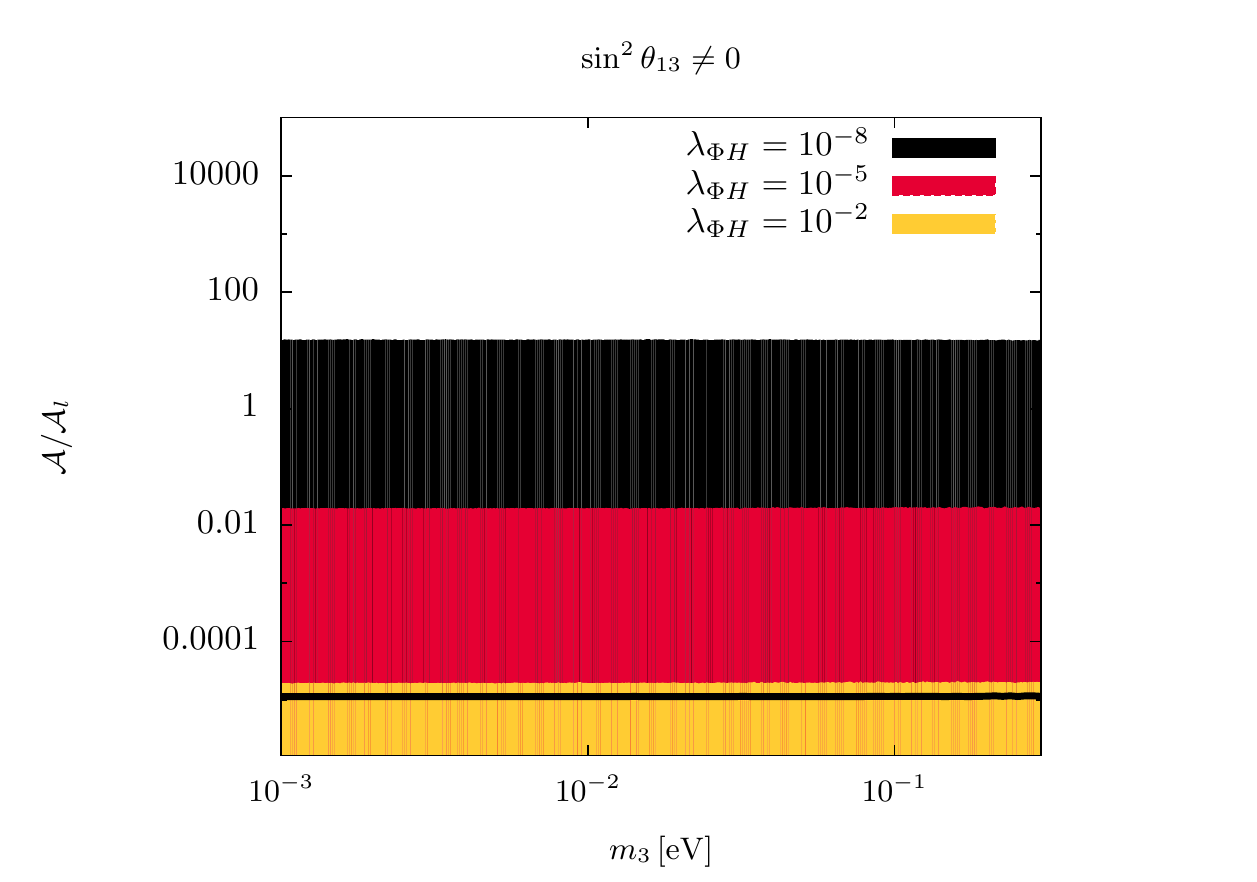}
\end{minipage}
\caption{The ratio of particle physics amplitudes for normal hierarchy (left
panel) and inverted hierarchy (right panel) as a function of the lightest neutrino mass for non-zero $\sin^2
\theta_{13}$ \cite{t2k}. The parameters used are given in the text. The colored
areas give the allowed regions using the current MEG limit for $\mu
\rightarrow e\gamma$. Recall that the standard effective mass, or
$\mathcal{A}_l$, is observable in the next few years for values of $m_1
\gs 0.3$ eV. The black and red lines give the upper or lower values of the correspondingly colored areas. Note that the areas for different values of $\lambda_{\Phi H}$ overlap. The lower red and yellow  lines are below the scale of the
axis. 
}\label{fig:tbmmixing1}
\end{figure}

\begin{figure}[ht]
 \centering
\begin{minipage}{0.45\linewidth}
\centering
\includegraphics[width=\textwidth]{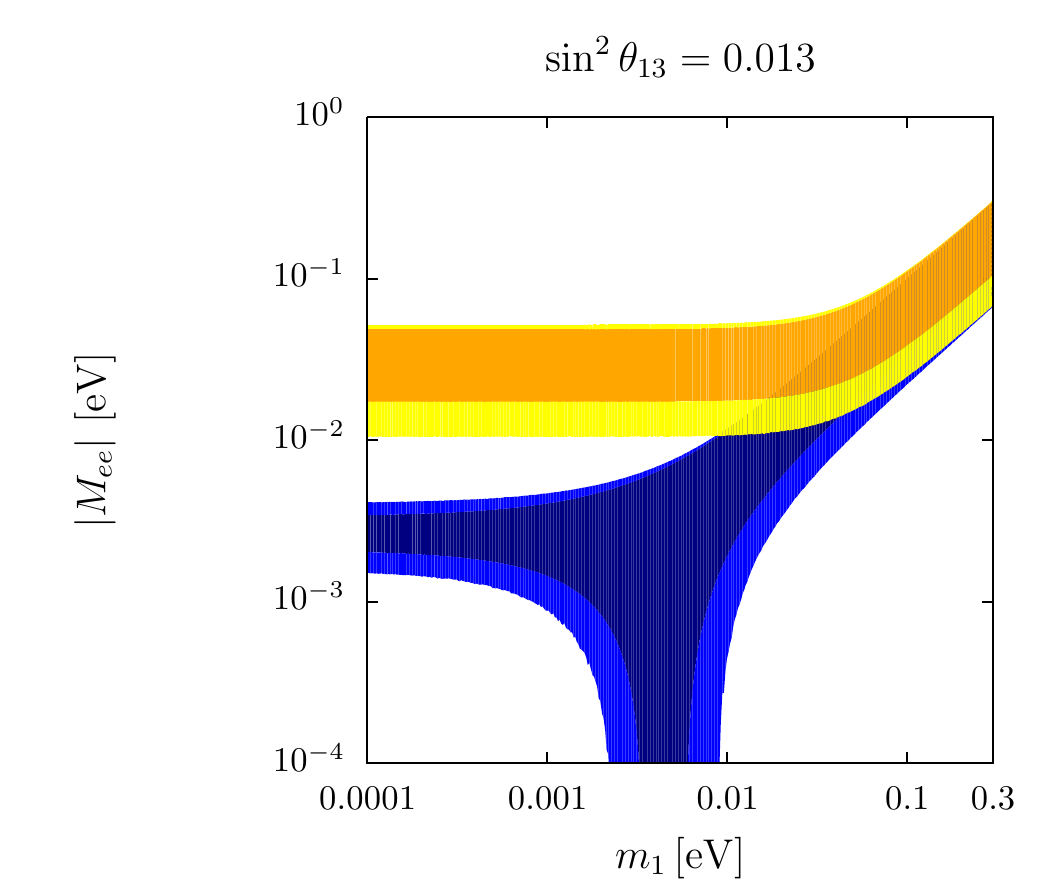}
\end{minipage}
\hspace{.1\linewidth}
\begin{minipage}{0.4\linewidth}
\centering
\includegraphics[width=\textwidth]{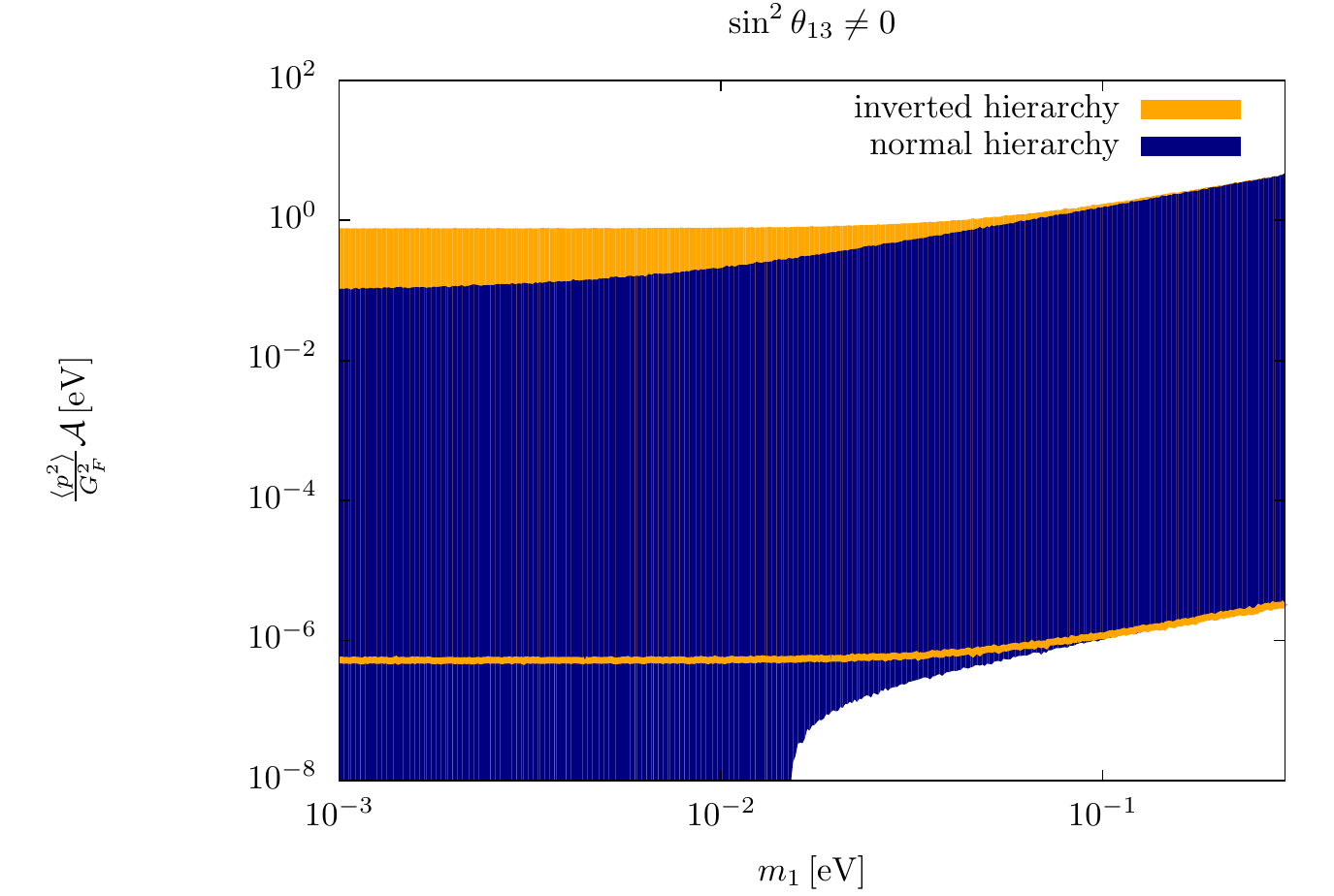}
\end{minipage}
\caption{Left: the usual plot \cite{vis-mee} of the effective mass against the
smallest neutrino mass. Right: the ``color effective mass'' 
$\frac{\langle p^2 \rangle}{G_F^2}\mathcal{A}$, which has to
be compared with the effective mass in the standard diagram for \obb,
as a function of the lightest neutrino mass. For the right plot, $\lambda_{\Phi H} = 10^{-8}$. The normal ordering is given in
blue and the inverted ordering in yellow. The darker areas are valid when the
oscillation parameters are fixed to their best-fit values (for better
visibility only best-fit values for the octet contribution are given), the brighter
areas for the $3\sigma$ ranges, and $\sin \theta_{13}$ has been fixed
to 0.013. In the right plot, the lower yellow line gives the minimum value for inverted hierarchy. Recall that the standard effective mass $M_{ee}$
is observable in the next few years for values of $m_1 \gs 0.3$ eV,
and the current limit on the half-life corresponds to about 1 eV for
$M_{ee}$, which roughly is also the limit for $\frac{\langle p^2 \rangle}{G_F^2}\mathcal{A}$. }\label{fig:mee}
\end{figure}

Focussing on the ratio between $\mathcal{A}$ and $\mathcal{A}_l$, and
therefore on the relative size of the direct and indirect contributions
to \obb, see Fig.~\ref{fig:tbmmixing1}. We have plotted there for the
normal and inverted mass ordering the ratio of the amplitudes as a
function of the smallest neutrino mass for different values of
$\lambda_{\Phi H}$. It is obvious that even for this 
simple example that the ratio of the amplitude can be very large or very small,
corresponding to the dominance of one of the contributions. 
The details of the figure are as follows: 
\begin{itemize}
\item
The mass of the color octet scalar $M_{\Phi}$ has been set to 2 TeV,
while the masses of the color octet fermions $M_{\Psi_i}$ have been
varied inside the interval $[ 0.9,1.1 ]$ TeV. No particular form of
hierarchy has been considered between the color octet fermions. The
random variation inside this interval mainly assures that there is no
exact degeneracy between the three fermions.
\item
The Yukawa coupling $y_{11}$ has been varied inside the interval $[0.001,1.0]$. Additionally, all phases in the PMNS matrix have been
varied in the interval $[0, 2\pi]$.
\item
The solar and atmospheric mass-squared differences, as well as
$\theta_{13}$ have been varied inside their presently allowed $3\sigma$ intervals \cite{Schwetz:2008er}. The typical momentum scale has been been set to $\langle p^2\rangle \simeq (100)^2\,
\rm{ MeV}^2$.
\item The differently colored regions in this figure correspond to different
$\lambda_{\Phi H}$ values as shown, all satisfying 
the MEG limit ${\rm Br} (\mu \to e\gamma)< 2.4 \times 10^{-12}$.
\item
Note that as $\lambda_{\Phi H}$ increases, the ratio ${\mathcal{A}}/{\mathcal{A}_l}$ decreases. The large increase in ${\mathcal{A}}/{\mathcal{A}_l}$ for low values of $m_1$ in normal hierarchy is an artifact 
of phase cancellation in $\mathcal{A}_l$. 
\end{itemize}
Another illustrative way to visualize the different contributions is
to define an ``effective mass'' for the direct octet
contribution. Noting that from the indirect amplitude the usual
effective mass $M_{ee}$ \cite{vis-mee, vis-lat} is obtained by multiplying $\mathcal{A}_l$ with $\langle
p^2\rangle/G_F^2$, we can define the ``color effective mass''  as
$\frac{\langle p^2\rangle}{G_F^2} \mathcal{A}$,  
where $\mathcal{A}$ is given in Eq.~(\ref{amp3a}). 
We can plot now both
the standard effective mass $M_{ee}$ and its analogous expression
$\frac{\langle p^2\rangle}{G_F^2} \mathcal{A}$ as a function of the lightest neutrino mass
eigenvalue. The two plots are given in Fig.~\ref{fig:mee}. We see that
the usual phenomenology can be significantly modified. For 
instance, in the inverted hierarchy (negligible $m_3$) one expects in the standard
case $M_{ee} \gs 0.05$ eV. The direct contribution from the octets
does approach 1 eV, and hence (for the simple example considered
here), can be used to cut in the parameter space of couplings and
masses. Note also that the predicted \obb~is very large even for normal hierarchy 
and almost comparable to that for inverted hierarchy, as pointed out earlier for 
the two octet case.

\section{Final Remarks and Conclusions \label{sum}}

There are two issues regarding our framework which we now address. 
The first one deals with
dropping the MFV hypothesis. Strong constraints on FCNC processes
exist, for example from scalar color octet exchange in $K^0$--$\bar{K^0}$
mixing, from $b \rightarrow s \gamma$ or the electric dipole moment of the
neutron, see e.g.~Ref.~\cite{Manohar:2006ga}. We note that the octet
contribution to neutrinoless double decay that we consider in this
work depends only on the coupling of the scalar octet $\Phi$ with an up- and a down-quark. One
can convince oneself that in all possible FCNC diagrams 
this coupling never appears on its own. For instance, in
$K^0$--$\bar{K^0}$ mixing diagrams or in $b \rightarrow s \gamma$ it
appears together with couplings involving 2nd and 3rd generation
quarks. Constraints coming from the electric dipole moment of the neutron
can be avoided by setting a possible phase to 
small values (in analogy to the SUSY CP problem). While this is not a
completely satisfying situation, we nevertheless note that in the
limit of only the coupling to up- and a down-quarks being non-zero, we
face no phenomenological problem. In addition, neutrinoless double
decay is the only place in which that coupling appears on its own and
hence it is the only place where it can directly be constrained. 

Another point is the strong hierarchy between $\lambda_{\Phi H}$ and
the Yukawa coupling of up- and down quarks. Radiative corrections might
spoil this hierarchy, for instance one might have diagrams in which
the quartic $\Phi^\dagger \Phi^\dagger H H $ coupling is mediated by quark
loops. However, in the same limit as above, namely only the coupling
of $\Phi$ to up- and a down-quarks being non-zero, this diagram is
suppressed heavily by $(m_{u,d}/v)^2$ and causes no problem. 

To sum up, in this work we have discussed 
neutrinoless double beta decay and lepton flavor violation in 
$\mu \to e\gamma$ for the so-called colored seesaw scenario. In this model, the color
octet scalars and fermions generate Majorana masses for the light
neutrinos via one-loop diagrams. Since these states have non-trivial charge
under $SU(3)_c$ and the electroweak gauge group, the same set of fields
can directly participate in neutrinoless double beta decay and lepton flavor violating processes. 

Studying only simple examples, we have already found interesting
features: it is for instance possible that the octet states saturate
the limits on both $\mu \to e \gamma$ and \onbb.
If the octet fermions
are degenerate in mass, then the contributions to \obb~from the octets and the
light neutrinos are both proportional to the effective mass $M_{ee}$, their
relative importance depending on the model parameters. 

It is conceptually interesting that the color octets imply a direct and
an indirect contribution to \onbb: the direct contribution is the one
considered here for the first time, namely the short-range exchange of octet
scalars and fermions. The indirect contribution is the standard, long-range one with the exchange of light Majorana neutrinos. These neutrinos are
generated at loop-level by the octets and it is interesting to compare
those two contributions. Extreme cases are easily possible, in the 
sense that both contributions can be either dominant or
negligible.

\vglue 0.8cm
\noindent
{\Large{\bf Acknowledgments}}\vglue 0.3cm
\noindent
MM would like to thank Francesco Vissani for carefully reading the manuscript and for his very useful comments. 
MD and WR would like to thank Adisorn Adulpravitchai for collaboration
at an early stage. MD is supported by the IMPRS-PTFS, WR by the ERC
under the Starting Grant MANITOP and by the Deutsche
Forschungsgemeinschaft in the Transregio 27. WR and SC acknowledge
support from the Indo-German DST-DFG program RO 2516/4-1. 
SC acknowledges support from the Neutrino Project
under the XIth plan of Harish-Chandra Research Institute.


\begin{thebibliography}{99}
\bibitem{revs}
 S.~M.~Bilenky, S.~T.~Petcov,
  Rev.\ Mod.\ Phys.\  {\bf 59}, 671 (1987);
%
M.~C.~Gonzalez-Garcia, Y.~Nir,
  Rev.\ Mod.\ Phys.\  {\bf 75}, 345-402 (2003)
  [arXiv:hep-ph/0202058].
%
R.~N.~Mohapatra, A.~Y.~Smirnov,
  Ann.\ Rev.\ Nucl.\ Part.\ Sci.\  {\bf 56} (2006)  569-628
  [arXiv:hep-ph/0603118];
  A.~Strumia and F.~Vissani,
  arXiv:hep-ph/0606054; 
  G.~Senjanovi\'c,
  Riv.\ Nuovo Cim.\ {\bf 034} (2011) 1-68.



\bibitem{epj}
H.~V.~Klapdor-Kleingrothaus, A.~Dietz, L.~Baudis, G.~Heusser, I.~V.~Krivosheina, S.~Kolb, B.~Majorovits, H.~Pas {\it et al.},
  Eur.\ Phys.\ J.\  {\bf A12 } (2001)  147-154 
  [arXiv:hep-ph/0103062].
  %
\bibitem{cuoricino+nemo}
 C.~Arnaboldi {\it et al.} [CUORICINO Collaboration],
  Phys.\ Rev.\  {\bf C78}, 035502 (2008) 
  [arXiv:0802.3439 [hep-ex]].
%
  J.~Argyriades {\it et al.} [NEMO Collaboration],
  Phys.\ Rev.\  {\bf C80}, 032501 (2009) 
  [arXiv:0810.0248 [hep-ex]].
%

\bibitem{Gerda}
 I.~Abt, M.~F.~Altmann, A.~Bakalyarov, I.~Barabanov, C.~Bauer, E.~Bellotti, S.~T.~Belyaev, L.~B.~Bezrukov {\it et al.},
  [arXiv:hep-ex/0404039]; S.~Schonert {\it et al.} [GERDA Collaboration],
  Nucl.\ Phys.\ Proc.\ Suppl.\  {\bf 145}, 242-245 (2005).
%
\bibitem{Cuore}
C.~Arnaboldi {\it et al.} [CUORE Collaboration],
 Nucl.\ Instrum.\ Meth.\  {\bf A518}, 775-798 (2004) 
  [arXiv:hep-ex/0212053].
%
\bibitem{MEG}
 J.~Adam {\it et al.} [MEG Collaboration],
  Phys.\ Rev.\ Lett.\  {\bf 107}, 171801 (2011) 
  [arXiv:1107.5547 [hep-ex]].
\bibitem{belle}
K.~Hayasaka {\it et al.} [Belle Collaboration],
  Phys.\ Lett.\ B\ {\bf 666}, 16  (2008)
  [arXiv:0705.0650 [hep-ex]].
%
\bibitem{future}
  R.~Arnold {\it et al.} [SuperNEMO Collaboration],
  Eur.\ Phys.\ J.\  {\bf C70}, 927-943 (2010)
  [arXiv:1005.1241 [hep-ex]];
%
  V.~E.~Guiseppe {\it et al.}  [Majorana Collaboration],
  IEEE Nucl.\ Sci.\ Symp.\ Conf.\ Rec.\  {\bf 2008} (2008) 1793
  [arXiv:0811.2446 [nucl-ex]].
%
F.~Ferroni,
  J.\ Phys.\ Conf.\ Ser.\  {\bf 293}, 012005 (2011)
  J.~W.~Beeman, F.~Bellini, L.~Cardani, N.~Casali, I.~Dafinei, S.~Di Domizio, F.~Ferroni and F.~Orio {\it et al.},
  Astropart.\ Phys.\  {\bf 35} (2012) 558
  [arXiv:1106.6286 [physics.ins-det]].
%
%
\bibitem{0nu2beta-old}
 G.~Racah,
  Nuovo Cim.\  {\bf 14}, 322-328 (1937);
  W.~H.~Furry,
  Phys.\ Rev.\  {\bf 56}, 1184-1193 (1939).
\bibitem{feinberg}
G.~Feinberg, M.~Goldhaber, Proc.\ Nat.\ Ac.\ Sci.\ USA\ {\bf 45}, 
1301  (1959);
  B.~Pontecorvo,
  Phys.\ Lett.\  {\bf B26}, 630-632 (1968).
  %
\bibitem{ms0nu2beta}
 R.~N.~Mohapatra,
 Phys.\ Rev.\  {\bf D34}, 3457-3461 (1986).
%
\bibitem{mwex}
 K.~S.~Babu, R.~N.~Mohapatra,
  Phys.\ Rev.\ Lett.\  {\bf 75}, 2276-2279 (1995)
  [arXiv:hep-ph/9506354].
%
\bibitem{pion-ex}
J.~D.~Vergados, Phys. Rev. D 25, 914 917 (1982);
  S.~Bergmann, H.~V.~Klapdor-Kleingrothaus, H.~Pas,
  Phys.\ Rev.\  {\bf D62}, 113002 (2000)
  [arXiv:hep-ph/0004048]; A.~Faessler, T.~Gutsche, S.~Kovalenko, F.~Simkovic,
   Phys.\ Rev.\  {\bf D77}, 113012 (2008)
   [arXiv:0710.3199 [hep-ph]].
%
\bibitem{hirsch}
  M.~Hirsch, H.~V.~Klapdor-Kleingrothaus, S.~G.~Kovalenko,
  Phys.\ Lett.\  {\bf B352}, 1-7 (1995) 
  [arXiv:hep-ph/9502315]; M.~Hirsch, H.~V.~Klapdor-Kleingrothaus, S.~G.~Kovalenko,
  Phys.\ Rev.\  {\bf D53}, 1329-1348 (1996)
  [arXiv:hep-ph/9502385];
   M.~Hirsch, H.~V.~Klapdor-Kleingrothaus, S.~G.~Kovalenko,
  Phys.\ Rev.\  {\bf D54}, 4207-4210 (1996)
  [arXiv:hep-ph/9603213];
  M.~Hirsch, H.~V.~Klapdor-Kleingrothaus, S.~G.~Kovalenko,
  Phys.\ Rev.\  {\bf D57}, 1947-1961 (1998)
  [arXiv:hep-ph/9707207];
   M.~Hirsch, J.~W.~F.~Valle,
  Nucl.\ Phys.\  {\bf B557}, 60-78 (1999)
  [arXiv:hep-ph/9812463].
%
\bibitem{allanach}
 B.~C.~Allanach, C.~H.~Kom, H.~Pas,
  Phys.\ Rev.\ Lett.\  {\bf 103}, 091801 (2009)
  [arXiv:0902.4697 [hep-ph]].
\bibitem{vogel}
 V.~Cirigliano, A.~Kurylov, M.~J.~Ramsey-Musolf, P.~Vogel,
  Phys.\ Rev.\  {\bf D70}, 075007 (2004)
  [arXiv:hep-ph/0404233];
 V.~Cirigliano, A.~Kurylov, M.~J.~Ramsey-Musolf, P.~Vogel,
  Phys.\ Rev.\ Lett.\  {\bf 93}, 231802 (2004)
  [arXiv:hep-ph/0406199].
\bibitem{choi}
K.~W.~Choi, K.~S.~Jeong, W.~Y.~Song,
  Phys.\ Rev.\  {\bf D66}, 093007 (2002)
  [arXiv:hep-ph/0207180].
\bibitem{tello}
V.~Tello, M.~Nemev\v sek, F.~Nesti, G.~Senjanovi\'c, F.~Vissani,
  Phys.\ Rev.\ Lett.\  {\bf 106}, 151801 (2011)
  [arXiv:1011.3522 [hep-ph]];
 M.~Nemevsek, F.~Nesti, G.~Senjanovic and V.~Tello,
  arXiv:1112.3061 [hep-ph].


\bibitem{Ibarra}
  A.~Ibarra, E.~Molinaro, S.~T.~Petcov,
  JHEP {\bf 1009}, 108 (2010)
  [arXiv:1007.2378 [hep-ph]];
 A.~Ibarra, E.~Molinaro, S.~T.~Petcov,
  [arXiv:1101.5778 [hep-ph]].
\bibitem{blennow}
  M.~Blennow, E.~Fernandez-Martinez, J.~Lopez-Pavon, J.~Menendez,
  JHEP {\bf 1007 } (2010)  096
  [arXiv:1005.3240 [hep-ph]].
%
\bibitem{msv}
 M.~Mitra, G.~Senjanovic and F.~Vissani,
  Nucl.\ Phys.\ B\ {\bf 856}, 26  (2012)
  [arXiv:1108.0004 [hep-ph]].
\bibitem{santamaria}
 F.~del Aguila, A.~Aparici, S.~Bhattacharya, A.~Santamaria and J.~Wudka,
  arXiv:1111.6960 [hep-ph].
%
\bibitem{Werner-rev} W.~Rodejohann,
  Int.\ J.\ Mod.\ Phys.\ E {\bf 20} (2011) 1833
  [arXiv:1106.1334 [hep-ph]].
%
\bibitem{ital-rev}
  J.~J.~Gomez-Cadenas, J.~Martin-Albo, M.~Mezzetto, F.~Monrabal and M.~Sorel,
  Riv.\ Nuovo Cim.\  {\bf 35} (2012) 29
  [arXiv:1109.5515 [hep-ex]].


\bibitem{seesaw}
 P.~Minkowski,
  Phys.\ Lett.\  {\bf B67}, 421 (1977).
 R.~N.~Mohapatra, G.~Senjanovi\'c,
  Phys.\ Rev.\ Lett.\  {\bf 44}, 912 (1980).
 T.~T.~Yanagida, in {\it Proceedings of the Workshop on the Unified Theory and the Baryon Number in the Universe} (O. Sawada and A. Sugamoto, eds.), KEK, Tsukuba, Japan, 1979, p.95.
 M.~Gell-Mann, P.~Ramond, R.~Slansky, Supergravity (P. van Nieuwenhuizen {\it et al.}, eds.), North Holland, Amsterdam, 1980.

\bibitem{FileviezPerez:2009ud}
  P.~Fileviez Perez and M.~B.~Wise,
  Phys.\ Rev.\ D {\bf 80} (2009) 053006
  [arXiv:0906.2950 [hep-ph]].
\bibitem{FileviezPerez:2010ch}
  P.~Fileviez Perez, T.~Han, S.~Spinner and M.~K.~Trenkel,
  JHEP {\bf 1101} (2011) 046
  [arXiv:1010.5802 [hep-ph]].

\bibitem{Liao:2009fm}
 Y.~Liao and J.~Y.~Liu,
  Phys.\ Rev.\  D {\bf 81} (2010) 013004
  [arXiv:0911.3711 [hep-ph]].


\bibitem{dim5}
 S.~Weinberg,
  Phys.\ Rev.\ Lett.\  {\bf 43}, 1566-1570 (1979);
   F.~Wilczek, A.~Zee,
  Phys.\ Rev.\ Lett.\  {\bf 43}, 1571-1573 (1979).
\bibitem{casas-Ibarra}
J.~A.~Casas, A.~Ibarra,
  Nucl.\ Phys.\  {\bf B618}, 171-204 (2001) 
  [arXiv:hep-ph/0103065].

\bibitem{LHC}
  G.~Aad {\it et al.}  [ATLAS Collaboration],
  Phys.\ Lett.\ B {\bf 708} (2012) 37
  [arXiv:1108.6311 [hep-ex]].

\bibitem{LHC1}
G.~Aad {\it et al.}  [ATLAS Collaboration],
 arXiv:1109.6572 [hep-ex].


\bibitem{Manohar:2006ga}
  A.~V.~Manohar and M.~B.~Wise,
  Phys.\ Rev.\ D {\bf 74} (2006) 035009
  [arXiv:hep-ph/0606172].

\bibitem{t2k}
  K.~Abe {\it et al.}  [T2K Collaboration],
  Phys.\ Rev.\ Lett.\  {\bf 107} (2011) 041801
  [arXiv:1106.2822 [hep-ex]].
\bibitem{Schwetz:2008er}
  T.~Schwetz, M.~A.~Tortola and J.~W.~F.~Valle,
  New J.\ Phys.\  {\bf 10} (2008) 113011
  [arXiv:0808.2016 [hep-ph]].


\bibitem{vis-mee}
 F.~Vissani,
  JHEP {\bf 9906}, 022 (1999)
  [arXiv:hep-ph/9906525].

\bibitem{vis-lat} 
  F.~Feruglio, A.~Strumia and F.~Vissani,
  Nucl.\ Phys.\ B {\bf 637}, 345 (2002)
  [Addendum-ibid.\ B {\bf 659}, 359 (2003)]
  [arXiv:hep-ph/0201291];
 A.~Strumia and F.~Vissani,
  Nucl.\ Phys.\ B {\bf 726}, 294 (2005)
  [arXiv:hep-ph/0503246].

\end{thebibliography}
\end{document}